\documentstyle[seceq,preprint,epsf]{jpsj}

\makeatletter
\def\lesssim{\mathrel{\mathpalette\vereq<}}
\def\gtrsim{\mathrel{\mathpalette\vereq>}}
\def\vereq#1#2{\lower3pt\vbox{\baselineskip1.5pt \lineskip1.5pt
\ialign{$\m@th#1\hfill##\hfil$\crcr#2\crcr\sim\crcr}}}
\makeatother

\def\runtitle{
    Transmission Probability for Interacting Electrons 
  Connected to Reservoirs
}

\def\runauthor{Akira {\sc Oguri}}

\title{ 
    Transmission Probability for Interacting Electrons \\
   Connected to Reservoirs
}

\author{Akira {\sc Oguri}}

\inst{
          Department of Material Science,
          Osaka City University, 
          Sumiyoshi-ku, Osaka 558-8585,
          Japan 
          \\
         and 
         \\
          Department of Mathematics, Imperial College, 
          180 Queen's Gate, London SW7 2BZ, UK
}

\recdate{May 14, 2001}

\abst{
Transport through small interacting systems 
connected to noninteracting leads is studied  
based on the Kubo formalism using a \'{E}liashberg theory of 
the analytic properties of the vertex part. 
The transmission probability, by which 
the conductance is expressed as  
 $g = (2e^2/h) \int {\rm d}\epsilon\, (- \partial f / \partial \epsilon) \, 
 {\cal T}(\epsilon)$, 
is introduced for interacting electrons. 
Here $f(\epsilon)$ is the Fermi function, and  
the transmission probability ${\cal T}(\epsilon)$ is defined 
in terms of a current vertex or 
a three-point correlation function.
We apply this formulation to a series of Anderson impurities    
of size $N$ ($=1,2,3,4$), 
and calculate ${\cal T}(\epsilon)$ using the order $U^2$ self-energy 
and current vertex which satisfy a generalized Ward identity.
The results show that ${\cal T}(\epsilon)$ has much information 
about the excitation spectrum: 
${\cal T}(\epsilon)$ has two broad peaks of 
the upper and lower Hubbard bands 
in addition to $N$ resonant peaks which 
have direct correspondence with the noninteracting spectrum.
The peak structures disappear at high temperatures. 
}

\kword{
transmission probability,
Kubo formula,
interacting electrons,
vertex correction,
analytic continuation 
}

\begin{document}
\sloppy
\maketitle

\section{Introduction}
\label{sec:Introduction}

Effects of electron correlation on 
transport through small interacting systems 
have been a subject of current interest. 
For instance, the Kondo effect\cite{Hewson} 
in quantum dots has been studied intensively, 
and recent experiments\cite{Goldharber-Gordon,Kouwenhoven,Simmel} 
have demonstrated a qualitative agreement 
with theoretical predictions.\cite{NL,GR,Kawabata}

To investigate transport properties of small systems,
it is essential to use theoretical approaches 
which are able to treat correctly both 
the interaction and quantum interference effects.
The Keldysh formalism\cite{Keldysh} 
has been applied to nonequilibrium systems.\cite{Caroli,MW} 
Caroli {\em et al\/} have shown for noninteracting electrons 
that the current is determined by single-particle excitations 
between the two chemical potentials 
of the left and right leads.\cite{Caroli} 
However, to our knowledge, 
the same feature has been shown to hold for interacting electrons   
only in a special case where the hopping matrix elements, 
which connects an interacting region at the center 
and the two leads, satisfy a certain condition.\cite{MW} 
Although this condition for the connectivity 
is not so important for a single Anderson impurity, 
it restricts the application of the formulation to 
interacting systems consisting of a number of quantum levels.
Our motivation of a series of works is based on a conjecture: 
as far as the linear response is concerned, 
the current through small interacting systems may be determined
by low-lying states near the Fermi energy. 
It has been clarified without the assumption of the connectivity  
that the conductance is written in terms of  
the Green's function at the Fermi energy $\epsilon=0$,
at zero temperature $T=0$.\cite{ao5,ao6} 
This is caused by a ground-state property of 
Fermi systems, i.e., the imaginary part of the self-energy 
vanishes at $\epsilon=0$, $T=0$.\cite{LangerAmbegaokar} 
Consequently, the contribution of the vertex corrections vanishes. 
The proof was provided based on the Kubo formalism  
making use of the diagrammatic analysis,\cite{YamadaYosida,Shiba} 
and it justifies an interpretation of the conductance 
in terms of the free quasi-particles.\cite{ao7,ao9}

The purpose of this paper is to generalize  
the formulation to finite temperatures. 
To this end, we have carried out 
the analytic continuation of the vertex part 
making use of a \'{E}liashberg theory of 
the analytic properties,  
which was originally proposed to the translational 
invariant Fermi systems\cite{Eliashberg}
and extended to lattice systems.\cite{YY_PAM} 
Compared to these examples, the situation we are considering 
is different in the following points: 
systems have no translational symmetry, 
and noninteracting leads connected to the interacting region 
keep the initial and final scattering states unrenormalized. 
One of the main results is that 
the conductance 
can be expressed as eq.\ (\ref{eq:cond1}) 
using the transmission probability ${\cal T}(\epsilon)$  
which is defined in terms of a current vertex 
eq.\ (\ref{eq:T_eff_sp}) or 
a three-point correlation function eq.\ (\ref{eq:T_eff2}). 
It can be also written in the
 Lehmann representation as eq.\ (\ref{eq:Lehmann_3}).

We apply the formulation to a series of Anderson impurities 
of size $N$ (=1,2,3,4).
For $N \geq 2$, the off-diagonal (non-local) elements 
of the self-energy and current vertex   
play important roles on the transport properties.\cite{ao9}
Therefore, it is essential to perform calculations
with a current-conserving approach. 
We calculate ${\cal T}(\epsilon)$ in an electron-hole symmetric case 
using the order $U^2$ self-energy and current vertex 
which satisfy a generalized Ward identity. 
At low temperatures, the value of the conductance 
is determined by the low energy part of ${\cal T}(\epsilon)$ because of 
the Fermi factor $-\partial f/ \partial \epsilon$ in eq.\ (\ref{eq:cond1}).
Nevertheless, the high energy part of ${\cal T}(\epsilon)$ also 
has much information about the excitation spectrum 
of interacting electrons: 
${\cal T}(\epsilon)$ has two broad peaks of 
the upper and lower Hubbard bands in addition 
to $N$ resonant peaks 
which have direct correspondence with the noninteracting spectrum.
At low temperatures, 
the peak structures near the Fermi energy depend on 
whether $N$ is even or odd: 
the Kondo peak is situated at $\epsilon=0$ for odd $N$
while it is a valley for even $N$.
The even-odd oscillation becomes small with increasing $T$  
and disappears at high temperatures $T \gtrsim T_K$,
where $T_K$ is the Kondo temperature.

In \S \ref{sec:CONDUCTANCE}, 
we describe the formulation, 
and introduce the transmission probability ${\cal T}(\epsilon)$ 
carrying out the analytic continuation of the vertex part.
In \S \ref{sec:WardIdentity},
the alternative expression of ${\cal T}(\epsilon)$ is given 
in terms of a three-point correlation function, and 
the current conservation is discussed in terms of 
a generalized Ward identity.
In \S \ref{sec:2nd_order}, 
we apply the formulation to a linear chain 
of the Anderson impurities.
Summary is given in \S \ref{sec:SUMMARY}. 
The Lehmann representation of ${\cal T}(\epsilon)$ is given 
in Appendix.

\section{Conductance and Transmission Probability}
\label{sec:CONDUCTANCE}

In this section
we describe a finite temperature formulation 
for the current through small interacting systems connected to reservoirs
 based on the linear response theory.
We perform the analytic continuation 
of the vertex part following \'{E}liashberg,\cite{Eliashberg}
and introduce the transmission probability for 
interacting electrons, ${\cal T}(\epsilon)$,   
by which the conductance can be expressed 
in a Landauer type form.\cite{Landauer,Buttiker,FisherLee}

\subsection{Model}
\label{subsec:model}

\begin{figure}
\centerline{ \vbox{ \epsfxsize=150mm \epsfbox {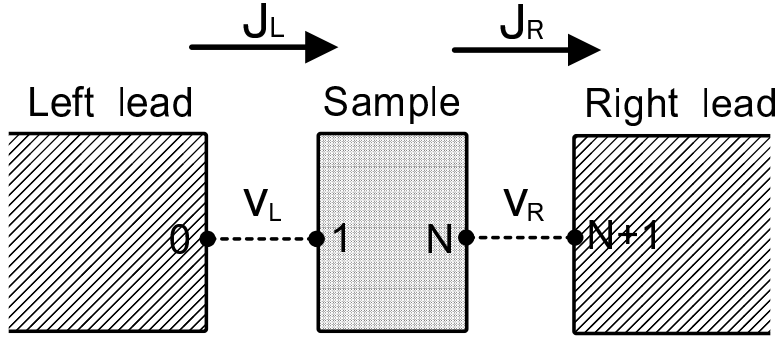}}
}
\vspace{-13cm}
\caption{ Schematic picture of the system.} 
\label{fig:single}
\end{figure}

We start with a system which consists of three regions; 
a finite central region ($C$) 
and two reservoirs on the left ($L$) and the right ($R$).
The central region consists of $N$ resonant levels, 
and the interaction  $U_{j_4 j_3; j_2 j_1}$ is 
switched on only for the electrons in this region.
We assume that each of the reservoirs is infinitely large 
and has a continuous energy spectrum.
The central region and the reservoirs are connected with  
the mixing matrix elements $v_L^{\phantom{\dagger}}$ 
and $v_R^{\phantom{\dagger}}$, as illustrated in Fig.\ \ref{fig:single}. 
The Hamiltonian is given by 
\begin{eqnarray}
{\cal H} \ 
    &=&  \ {\cal H}_L + {\cal H}_R  + {\cal H}_C^0 + {\cal H}_C^{int}
            + {\cal H}_{mix}     
\label{eq:H}            
\;, \\
{\cal H}_L &=&  \sum_{ij\in L} \sum_{\sigma} 
        \left(\,-t_{ij}^L - \mu\, \delta_{ij} \,\right)
            c^{\dagger}_{i \sigma} c^{\phantom{\dagger}}_{j \sigma}      
             \;, 
\label{eq:H_L}           
\\ 
{\cal H}_R  &=&  \sum_{ij\in R} \sum_{\sigma} 
        \left(\,-t_{ij}^R - \mu\, \delta_{ij} \,\right)
  c^{\dagger}_{i \sigma} c^{\phantom{\dagger}}_{j \sigma}        
      \;, 
\label{eq:H_R}      
\\      
  {\cal H}_C^{0} &=&   \sum_{ij\in C} \sum_{\sigma} 
        \left(\,-t_{ij}^C - \mu\, \delta_{ij} \,\right)  
   c^{\dagger}_{i \sigma} c^{\phantom{\dagger}}_{j \sigma}        
  \;, \\   
    {\cal H}_C^{int} &=&     
{1 \over 2} \sum_{\{j\} \in C}\sum_{\sigma \sigma'}    
    U_{j_4 j_3; j_2 j_1}\,  
 c^{\dagger}_{j_4 \sigma} c^{\dagger}_{j_3 \sigma'}    
 c^{\phantom{\dagger}}_{j_2 \sigma'} c^{\phantom{\dagger}}_{j_1 \sigma} 
  \;,  
\label{eq:H_int}  
\\
  {\cal H}_{mix} &=& 
-  \sum_{\sigma} v_L^{\phantom{\dagger}} \left(\,    
             c^{\dagger}_{1 \sigma} c^{\phantom{\dagger}}_{0 \sigma}
             +              
          c^{\dagger}_{0 \sigma} c^{\phantom{\dagger}}_{1 \sigma}        
                \,\right)
-  \sum_{\sigma} v_R^{\phantom{\dagger}} \left(\,            
             c^{\dagger}_{N+1 \sigma} c^{\phantom{\dagger}}_{N \sigma}
           + c^{\dagger}_{N \sigma} c^{\phantom{\dagger}}_{N+1 \sigma}                      \,\right)     
  .            
\label{eq:H_mix}  
\end{eqnarray}
Here $c^{\dagger}_{j \sigma}$ 
 ($c^{\phantom{\dagger}}_{j \sigma}$) creates (destroys) 
an electron with spin $\sigma$ at site $j$,  
and $\mu$ is the chemical potential.
$t_{ij}^{L}$, $t_{ij}^{R}$, and $t_{ij}^{C}$ are 
the intra-region hopping matrix elements
in each regions $L$, $R$, and $C$, respectively. 
The labels $1$, $2$, $\ldots$, $N$ are assigned to 
the sites in the central region.
Specifically, the label $1$ ($N$) is assigned to the site 
at the interface on the left (right), 
and the label $0$ ($N+1$) is assigned to the site 
at the reservoir-side of the left (right) interface.
We assume that the hopping matrix elements are real,
and the interaction has the time-reversal symmetry:  
$U_{4 3; 2 1}$ is real 
and $U_{4 3; 2 1}=U_{3 4; 1 2}=U_{1 2; 3 4 }=U_{4 2; 3 1}=U_{1 3; 2 4}$. 
We will be using units $\hbar=1$ unless otherwise noted.

The single-particle Green's function is defined by 
\begin{equation} 
G_{jj'}({\rm i}\varepsilon _m) 
 =  
-    \int_0^{\beta} \! {\rm d}\tau 
   \left \langle  T_{\tau} \,  
   c^{\phantom{\dagger}}_{j \sigma} (\tau) \, c^{\dagger}_{j' \sigma} (0)     
           \right \rangle  \, {\rm e}^{{\rm i}\, \varepsilon_m \tau} .  
\label{eq:G_Matsubara}                  
\end{equation} 
Here $\beta= 1/T$, $\varepsilon_m = (2m+1)\pi/\beta$, 
$c_{j \sigma}(\tau) = {\rm e}^{\tau  {\cal H}} c_{j \sigma} 
{\rm e}^{- \tau  {\cal H}}$,
and $\langle \cdots \rangle$ denotes the thermal average 
$\mbox{Tr} \left[ \, {\rm e}^{-\beta  {\cal H} }\, {\cdots}
\,\right]/\mbox{Tr} \, {\rm e}^{-\beta  {\cal H} }$.
The spin index has been omitted from the left-hand side 
of eq.\ (\ref{eq:G_Matsubara}) 
assuming the expectation value to be 
independent of whether spin is up or down.
Since the interaction is switched on only 
for the electrons in the central region,
the Dyson equation can be written as
\begin{equation} 
  G_{ij}(z)    =   G^0_{ij}(z) 
    + \sum_{i'j' \in C}\,G^0_{ii'}(z)\,  \Sigma_{i'j'}(z)  
   \, G_{j'j}(z) \;.    
  \label{eq:Dyson}   
\end{equation}   
Here $G^0_{ij}(z)$ is the unperturbed Green's function 
corresponding to the noninteracting Hamiltonian ${\cal H}^0
 \equiv   {\cal H}_L + {\cal H}_R  
   + {\cal H}_C^0 + {\cal H}_{mix}$.
The summations with respect to $i'$ and $j'$ run over
the sites in the central region, 
and $\Sigma_{i'j'}(z)$ is the self-energy 
due to the interaction ${\cal H}_C^{int}$. 
Owing to the time-reversal symmetry of ${\cal H}$, 
these functions are symmetric against the interchange
of the site indices: 
 $G_{ij}(z) =  G_{ji}(z)$ and $\Sigma_{ij}(z) = \Sigma_{ji}(z)$. 
Note that at $T=0$ the single-particle excitation 
at the Fermi energy $z= {\rm i}0^+$ does not decay, i.e.,
 $\mbox{Im}\,\Sigma_{ij}^+ (0)=0$.\cite{LangerAmbegaokar}  
We will treat $z$ as a complex variable,
and use the symbol $+$ ($-$) in the superscript 
as a label for the retarded (advanced) function:  
$\Sigma_{ij}^{\pm}(\epsilon) \equiv \Sigma_{ij}(\epsilon \pm {\rm i}0^+)$.

\subsection{Analytic continuation of the vertex part}
\label{subsec:AnalyticContinuation}

We next consider the conductance based on the Kubo formalism.
The conductance $g$ is given by 
the $\omega$-linear imaginary part of 
a current-current correlation function: 
\begin{eqnarray}
 g \  &=& \ e^2 \, \lim_{\omega \to 0}
     { K_{\alpha\alpha'}(\omega+{\rm i}0^+) 
      - K_{\alpha\alpha'}({\rm i}0^+) \over {\rm i}\, \omega } 
\label{eq:Kubo} \;, 
\\
K_{\alpha\alpha'}({\rm i}\nu_l) &=&   \int_0^{\beta} \! {\rm d}\tau 
\left\langle T_{\tau}\, J_{\alpha}(\tau) J_{\alpha'}(0) \right\rangle 
      \, {\rm e}^{{\rm i}\, \nu_l \tau}. 
\end{eqnarray}
Here $\nu_l = 2\pi l/\beta$ is the Matsubara frequency, 
and the retarded function 
can be obtained through the analytic continuation 
$K_{\alpha\alpha'}(\omega+{\rm i}0^+) \equiv  
\left. K_{\alpha\alpha'}({\rm i}\nu_l)\right|_{{\rm i}\nu_l 
\to \omega + {\rm i}0^+}$.
The current operator $J_{\alpha}$, for $\alpha = L$ or $R$, 
is defined by 
\begin{eqnarray}
 J_L  &=& {\rm i}\,     
\sum_{\sigma}
                  v_{L}^{\phantom{\dagger}} 
           \left(\, 
            c^{\dagger}_{1 \sigma} 
            c^{\phantom{\dagger}}_{0 \sigma}  
          - c^{\dagger}_{0 \sigma} 
            c^{\phantom{\dagger}}_{1 \sigma}
\, \right) 
\label{eq:J_L}
,\\
 J_R  &=& {\rm i}\,     
\sum_{\sigma}
         v_{R}^{\phantom{\dagger}}
  \left(\, 
           c^{\dagger}_{N+1 \sigma} 
           c^{\phantom{\dagger}}_{N  \sigma} 
        -  c^{\dagger}_{N \sigma} 
           c^{\phantom{\dagger}}_{N+1  \sigma} 
\, \right) 
\label{eq:J_R} .
\end{eqnarray}
Here $J_L$ is the current flowing into the sample 
from the left lead,  and 
$J_R$ is the current flowing out to the right lead from the sample.
These currents and total charge in the sample, 
$\rho_C^{\phantom{0}} =      
\sum_{j\in C, \sigma} 
 c^{\dagger}_{j \sigma} c^{\phantom{\dagger}}_{j \sigma}$,
satisfy the equation of continuity  
${\partial \rho_C^{\phantom{0}}/ \partial t} + J_R - J_L = 0$.
Due to this property, 
the value of $g$ given by eq.\ (\ref{eq:Kubo}) is 
independent of $\alpha$ and $\alpha'$.\cite{FisherLee}
Note that  $K_{\alpha'\alpha}(z)  =  K_{\alpha\alpha'}(z)$ 
because of the time-reversal symmetry of ${\cal H}$.

\begin{figure}
\vspace{-2cm}
\rule{3cm}{0cm}
\centerline{ \vbox{ \epsfxsize=100mm \epsfbox {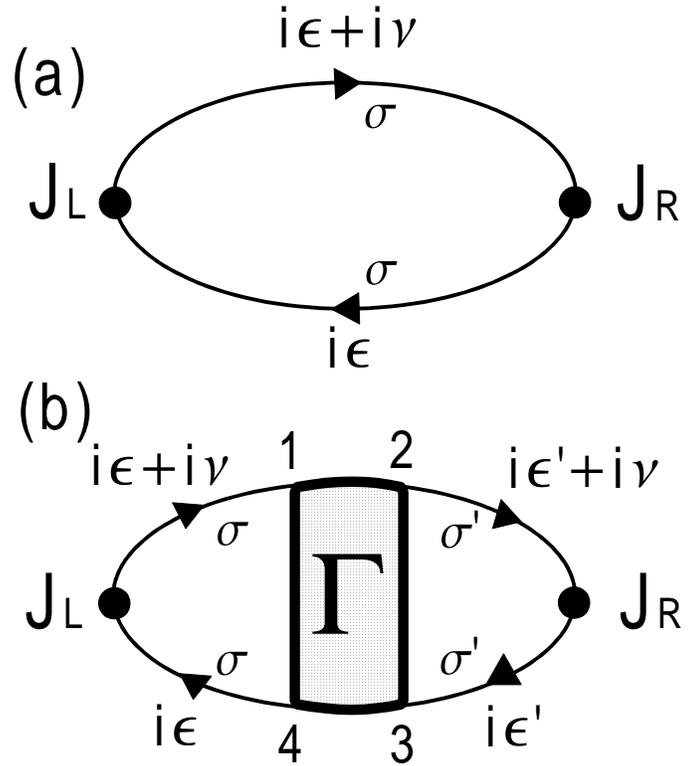}}
}
\vspace{-2cm}
\caption{Diagrams for $K_{RL}({\rm i}\nu)$. 
The shaded region represents the vertex part 
$\Gamma^{\sigma \sigma '}_{1 2;\, 3 4} 
({\rm i}\varepsilon , {\rm i}\varepsilon '; {\rm i}\nu )$.}
\label{fig:diagram1}
\end{figure}

\begin{figure}
\rule{4.5cm}{0cm}
\centerline{ \vbox{ \epsfxsize=80mm \epsfbox {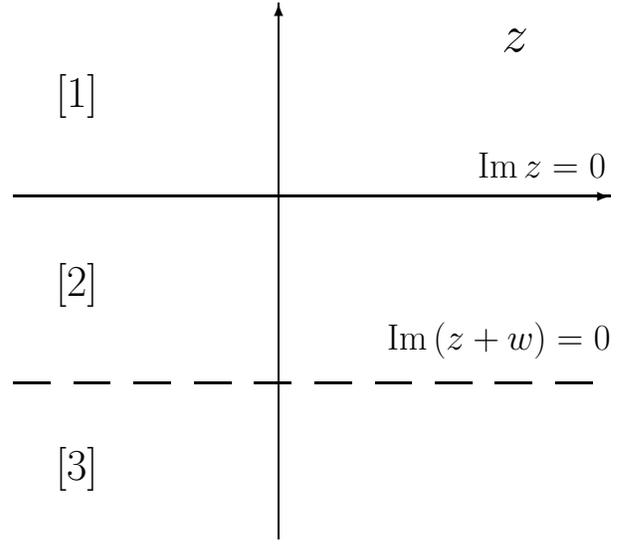}}
}
\caption{Three analytic regions of the current vertex
$\Lambda_{R;j_4j_1}(z, z+w)$.
}
\vspace{-1cm}
\label{fig:current_analytic}
\end{figure}

\begin{figure}
\rule{4.5cm}{0cm}
\centerline{ \vbox{ \epsfxsize=80mm \epsfbox {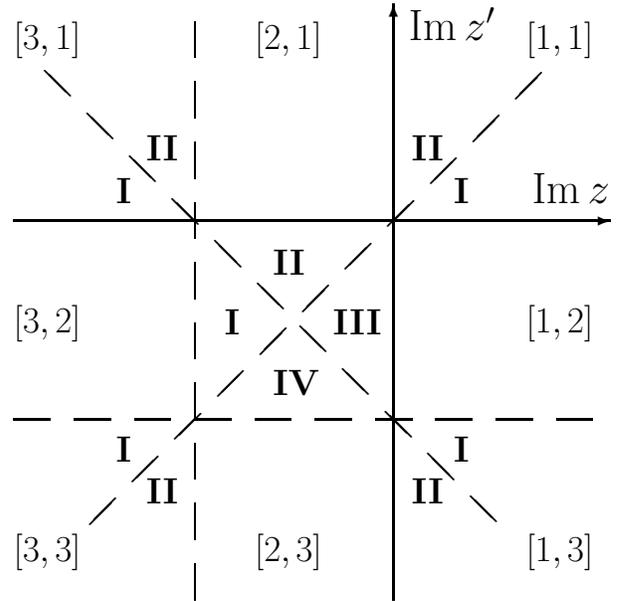}}
}
\caption{Analytic regions of the vertex part
 $\Gamma^{\sigma \sigma '}_{j_1 j_2;\, j_3 j_4}(z, z'; w )$.
}
\label{fig:vertex_analytic}
\end{figure}

We calculate the $\omega$-linear part of $K_{\alpha \alpha'}(z)$ 
taking $\alpha$ and $\alpha'$ to be $R$ and $L$.
The function $K_{RL}({\rm i}\nu)$ can be separated into the two terms 
as shown in Fig.\ \ref{fig:diagram1}: 
\begin{eqnarray}
K_{RL}({\rm i}\nu)  &=& 
\  K_{RL}^{(a)}({\rm i}\nu) \ + \ K_{RL}^{(b)}({\rm i}\nu) \; , 
\label{eq:K_nu} \\
K_{RL}^{(a)}({\rm i}\nu) &=&   -  
\, ({\rm i}\, v_{L}^{\phantom{\dagger}})\, 
({\rm i}\, v_{R}^{\phantom{\dagger}})
\  {1\over\beta}
\sum_{{\rm i}\varepsilon } 
\sum_{\sigma} 
\nonumber \\
& & \times \Bigl [ \, 
  G_{N,1}({\rm i}\varepsilon  + {\rm i}\nu)\, G_{0,N+1}({\rm i}\varepsilon ) 
 + G_{N+1,0}({\rm i}\varepsilon  + {\rm i}\nu)\, G_{1,N}({\rm i}\varepsilon ) 
\nonumber \\
& & \quad 
- G_{N+1,1}({\rm i}\varepsilon  + {\rm i}\nu)\, G_{0,N}({\rm i}\varepsilon ) 
- G_{N,0}({\rm i}\varepsilon  + {\rm i}\nu)\, G_{1,N+1}({\rm i}\varepsilon ) 
\, \Bigr ] \;,  
\label{eq:K_nu_a}
\\ 
K_{RL}^{(b)}({\rm i}\nu) &=&  -    
\, ({\rm i}\, v_{L}^{\phantom{\dagger}})\, 
({\rm i}\, v_{R}^{\phantom{\dagger}})
\ {1 \over \beta^2}
\sum_{{\rm i}\varepsilon , {\rm i}\varepsilon '} 
\sum_{\sigma \sigma '} 
\sum_{ \{j\} \in C}
\nonumber \\
& & \times \Bigl [ \, 
 G_{j_1,1}({\rm i}\varepsilon  + {\rm i}\nu)\, G_{0,j_4}({\rm i}\varepsilon ) 
- G_{j_1,0}({\rm i}\varepsilon  + {\rm i}\nu)\, G_{1,j_4}({\rm i}\varepsilon ) 
\, \Bigr ] 
\nonumber \\
& & \times \,\Gamma^{\sigma \sigma '}_{j_1 j_2;\, j_3 j_4}
({\rm i}\varepsilon , {\rm i}\varepsilon '; {\rm i}\nu) 
\nonumber \\
& & \times \Bigl [ \, 
 G_{N,j_2}({\rm i}\varepsilon ' + {\rm i}\nu)\, 
 G_{j_3,N+1}({\rm i}\varepsilon ')
- G_{N+1,j_2}({\rm i}\varepsilon ' + {\rm i}\nu)\, 
G_{j_3,N}({\rm i}\varepsilon ')
\, \Bigr ] 
\;.
\label{eq:K_nu_b}
\end{eqnarray}
Here
$\Gamma^{\sigma \sigma '}_{j_1 j_2;\, j_3 j_4}
({\rm i}\varepsilon , {\rm i}\varepsilon 
'; {\rm i}\nu)$
is the total vertex part due to ${\cal H}_C^{int}$, and
the summation with respect to $\{j\}$ 
runs over the sites in the central region  $j_1, \ldots, j_4 \in C$.
Since the leads are assumed to be noninteracting,
the Green's function satisfies the following relations 
at the two interfaces:
\begin{eqnarray}
  \left \{
  \begin{array}{ll} 
   G_{0, j}(z) \phantom{_{+1}}
  =  -
  \mbox{\sl g}_L(z)\, v_{L}^{\phantom{\dagger}} G_{1, j}(z)
   &\quad \mbox{for}\quad  1\leq j \leq N+1
   \\
   G_{j, N+1}(z) 
     =  - 
     G_{j, N}(z)\, v_{R}^{\phantom{\dagger}}\,\mbox{\sl g}_R(z)
      &\quad \mbox{for}\quad  0 \leq j \leq N\\
 \end{array}
  \right.
   \;\;.
\label{eq:rec_lead} 
\end{eqnarray}
Here $\mbox{\sl g}_L(z)$  and $\mbox{\sl g}_R(z)$ are 
the local Green's functions at the interfaces 
of the isolated leads, $0$ and $N+1$, respectively.
Using these relations,  
$K_{RL}({\rm i}\nu)$ can be rewritten 
in terms of the Green's functions in the central region as
\begin{eqnarray}
K_{RL}({\rm i}\nu) &=&               
 \ {1 \over \beta}
\sum_{{\rm i}\varepsilon } 
\sum_{\sigma} 
\sum_{ j_1j_4 \in C} 
\lambda_L({\rm i}\varepsilon , {\rm i}\varepsilon  + {\rm i}\nu)\,
G_{1 j_4}({\rm i}\varepsilon ) 
\, \Lambda_{R;j_4j_1}({\rm i}\varepsilon , {\rm i}\varepsilon  + {\rm i}\nu)
\,G_{j_1 1}({\rm i}\varepsilon +{\rm i}\nu) 
\;. 
\label{eq:K_nu_2}
\end{eqnarray}
Here the bare and full current vertices are defined by
\begin{eqnarray}
\lambda_L({\rm i}\varepsilon , {\rm i}\varepsilon  + {\rm i}\nu)
&=& -{\rm i}\, v_L^{2} 
\left[\, \mbox{\sl g}_L({\rm i}\varepsilon  + {\rm i}\nu) 
- \mbox{\sl g}_L({\rm i}\varepsilon )\,\right]
\;,
\label{eq:lambda_L0} 
\\
\lambda_R({\rm i}\varepsilon , {\rm i}\varepsilon  + {\rm i}\nu)
&=& \ {\rm i}\, v_R^{2} 
\left[\, \mbox{\sl g}_R({\rm i}\varepsilon  + {\rm i}\nu) 
- \mbox{\sl g}_R({\rm i}\varepsilon )\,\right]
\;,
\label{eq:lambda_R0} 
\\
\Lambda_{R;j_4j_1}({\rm i}\varepsilon, {\rm i}\varepsilon  + {\rm i}\nu) &=& 
\,
\lambda_R({\rm i}\varepsilon, {\rm i}\varepsilon  + {\rm i}\nu)
\, \delta_{N j_4}  \delta_{N j_1}
\, + \, P_{j_4j_1}({\rm i}\varepsilon, {\rm i}\varepsilon  + {\rm i}\nu) 
\;,
\label{eq:Lambda_R} 
\\
P_{j_4j_1}({\rm i}\varepsilon , {\rm i}\varepsilon  + {\rm i}\nu) 
&=& \, 
{1 \over \beta}
\sum_{{\rm i}\varepsilon '} 
\sum_{\sigma '} 
\sum_{ j_2 j_3 \in C}
\,\Gamma^{\sigma \sigma '}_{j_1 j_2;\, j_3 j_4}
({\rm i}\varepsilon , {\rm i}\varepsilon '; {\rm i}\nu) 
\nonumber \\  
& & \times 
\, G_{j_3 N}({\rm i}\varepsilon ') 
\,\lambda_R({\rm i}\varepsilon ', {\rm i}\varepsilon ' + {\rm i}\nu )  
\, G_{N j_2}({\rm i}\varepsilon ' + {\rm i}\nu)
\;.
\label{eq:P_M} 
\end{eqnarray}
The spin index has been omitted from the left-hand side of 
eq.\ (\ref{eq:P_M}) since it is independent of $\sigma$ 
due to the rotational symmetry of the spin space.
The $\omega$-linear part of the retarded function 
$K_{RL}(\omega + i 0^+ )$ can be obtained from eq.\ (\ref{eq:K_nu_2}) 
by carrying out the analytic continuation:
\begin{eqnarray}
K_{RL}(\omega + i 0^+ ) 
&=&                                          
\sum_{\sigma} \sum_{ j_1 j_4 \in C}
\biggl \{ 
\nonumber \\
& & - \int_{-\infty}^{\infty} {{\rm d}\epsilon \over 2 \pi {\rm i}}\ 
f(\epsilon)
\ \lambda_L^{[1]}(\epsilon,\epsilon + \omega) 
\, G_{1 j_4}^+(\epsilon) 
\, \Lambda_{R;j_4j_1}^{[1]}(\epsilon,\epsilon + \omega) 
\, G_{j_1 1}^+(\epsilon + \omega)
\nonumber \\
& &- \int_{-\infty}^{\infty} {{\rm d}\epsilon \over 2 \pi {\rm i}} \,
\Bigl[\, f(\epsilon+\omega) - f(\epsilon) \,\Bigr]
\ \lambda_L^{[2]}(\epsilon,\epsilon + \omega)
\, G_{1 j_4}^-(\epsilon) 
\, \Lambda_{R;j_4j_1}^{[2]}(\epsilon,\epsilon + \omega) 
\, G_{j_1 1}^+(\epsilon + \omega)
\nonumber \\
& &+ \int_{-\infty}^{\infty} {{\rm d}\epsilon \over 2 \pi {\rm i}}\ 
f(\epsilon+\omega)
\ \lambda_L^{[3]}(\epsilon,\epsilon + \omega) 
\, G_{1 j_4}^-(\epsilon) 
\, \Lambda_{R;j_4j_1}^{[3]}(\epsilon,\epsilon + \omega) 
\, G_{j_1 1}^-(\epsilon + \omega)
\, \biggr \} .
\label{eq:K_ret} 
\end{eqnarray}
Here $f(\epsilon) = [{\rm e}^{\beta \epsilon}+1]^{-1}$.
The superscript [k] for $k = 1, 2, 3$ is introduced 
for the current vertices in order to specify 
the three analytic regions of $\Lambda_{R;j_4j_1}(z,z + w)$ 
in the complex $z$-plane,  
which are separated by the two lines, 
$\mbox{Im}\, (z)=0$ and $\mbox{Im}\, (z+w)=0$,  
as shown in Fig \ref{fig:current_analytic}.
The explicit form of the bare current vertex is given by 
\begin{equation}
    \lambda_{\alpha}^{[k]}(\epsilon, \epsilon +\omega) \propto
 \left\{ 
   \begin{array}{ll}
 \,\mbox{\sl g}_{\alpha}^+(\epsilon + \omega) - 
           \mbox{\sl g}_{\alpha}^+(\epsilon) 
           &\quad \mbox{for} \quad k=1     
            \\
   \,  \mbox{\sl g}_{\alpha}^+(\epsilon + \omega) - 
     \mbox{\sl g}_{\alpha}^-(\epsilon) 
           &\quad \mbox{for} \quad k=2     
           \\
    \,     \mbox{\sl g}_{\alpha}^-(\epsilon + \omega) - 
       \mbox{\sl g}_{\alpha}^-(\epsilon)
           &\quad \mbox{for} \quad k=3      
    \end{array} \right.
 \;,
\end{equation}
for $\alpha =L, R$. 
Thus, in the limit of $\omega \to 0$, 
$\lambda_{\alpha}^{[1]}(\epsilon, \epsilon) =0$ and 
$\lambda_{\alpha}^{[3]}(\epsilon, \epsilon) =0$.
Furthermore, 
$\lambda_R^{[2]}(\epsilon, \epsilon)
= 2\, \Gamma_R(\epsilon)$ and  $\lambda_L^{[2]}(\epsilon, \epsilon)
= -2\, \Gamma_L(\epsilon)$ with  
$\Gamma_{\alpha}(\epsilon) = -\, \mbox{Im} 
\left[ v_{\alpha}^2 \, \mbox{\sl g}_{\alpha}^{+}(\epsilon)  \right]$. 
In order to calculate the $\omega$-linear part of eq.\ (\ref{eq:K_ret}),  
we next consider the behavior 
of $P_{j_4j_1}^{[k]}(\epsilon, \epsilon +\omega)$ for small $\omega$ 
 carrying out the analytic continuation 
of $P_{j_4j_1}({\rm i}\varepsilon , {\rm i}\varepsilon  + {\rm i}\nu)$
using eq.\ (\ref{eq:P_M}).
It can be done following the \'{E}liashberg theory
of the analytic properties of the vertex part.\cite{Eliashberg} 
As a function of the complex variables, 
$\Gamma^{\sigma \sigma '}_{j_1 j_2;\, j_3 j_4}(z, z'; w)$
has singularities along the lines shown in Fig.\ \ref{fig:vertex_analytic}:
the vertical lines $\mbox{Im}\, (z)=0$ and $\mbox{Im}\, (z+w)=0$, 
the horizontal lines $\mbox{Im}\, (z')=0$ and $\mbox{Im}\, (z'+w)=0$, 
and the diagonal lines $\mbox{Im}\, (z-z')=0$ and $\mbox{Im}\, (z+z'+w)=0$. 
These lines divide the plane into 16 regions,
each of which corresponds to a function of 
$\Gamma^{\sigma \sigma '}_{j_1 j_2;\, j_3 j_4}(z, z'; w)$
which is analytic in that region in any of its arguments.
The rectangular regions in Fig.\ \ref{fig:vertex_analytic} are numbered by 
indices $[k,k']$ for  $k,k'=1,2,3$. 
Some of these regions are divided into parts by the diagonal lines,
and these parts are denoted by Roman numbers.
We will use this notation also for the vertex part 
in eqs.\ (\ref{eq:L11})--(\ref{eq:L33}).
Rewriting the summation over the Matsubara frequency ${\rm i}\varepsilon '$
in eq.(\ref{eq:P_M}) in terms of a contour integral along 
these lines  and then performing the analytic continuation,
$P_{j_4j_1}^{[k]}(\epsilon, \epsilon + \omega)$ is obtained as
\begin{eqnarray}
P_{j_4j_1}^{[k]}(\epsilon,\epsilon +\omega) 
 &=&  
\sum_{\sigma'} \sum_{ j_2 j_3 \in C}
\biggl \{\, 
\nonumber \\
& &  
\phantom{+} \int_{-\infty}^{\infty} {{\rm d}\epsilon' \over 2\pi {\rm i}}
\ {\cal L}^{\sigma \sigma' \, [k,1]}_{j_1 j_2;\, j_3 j_4}
(\epsilon,  \epsilon'; \omega) 
\ G_{j_3 N}^+(\epsilon') 
\, \lambda_{R}^{[1]}(\epsilon',\epsilon' + \omega) 
\, G_{N j_2}^+(\epsilon' + \omega)
\nonumber \\
& &  
+ \int_{-\infty}^{\infty} {{\rm d}\epsilon' \over 2\pi {\rm i}}
\ {\cal L}^{\sigma \sigma' \, [k,2]}_{j_1 j_2;\, j_3 j_4}
(\epsilon,  \epsilon'; \omega) 
\ G_{j_3 N}^-(\epsilon') 
\, \lambda_{R}^{[2]}(\epsilon',\epsilon' + \omega) 
\, G_{N j_2}^+(\epsilon' + \omega)
\nonumber \\
& &  
+ \int_{-\infty}^{\infty} {{\rm d}\epsilon' \over 2\pi {\rm i}}
\ {\cal L}^{\sigma \sigma' \, [k,3]}_{j_1 j_2;\, j_3 j_4}
(\epsilon,  \epsilon'; \omega) 
\  G_{j_3 N}^-(\epsilon') 
\, \lambda_{R}^{[3]}(\epsilon',\epsilon' + \omega) 
\, G_{N j_2}^-(\epsilon' + \omega)
\,\biggr \} \;.
\label{eq:P_ret1} 
\end{eqnarray}
Here 
${\cal L}^{\sigma \sigma' \,[k,k']}_{j_1 j_2;\, j_3 j_4}
(\epsilon,  \epsilon'; \omega)$ is defined by
\begin{eqnarray}
{\cal L}^{\sigma \sigma' \, [11]}_{j_1 j_2;\, j_3 j_4}
(\epsilon,  \epsilon'; \omega) 
&=& 
{\cal P} b(\epsilon'-\epsilon) 
\left[\, 
\Gamma^{\sigma \sigma'[11;\,\mbox{\scriptsize II}]}_{j_1 j_2;\, j_3 j_4}
(\epsilon,  \epsilon'; \omega) -
\Gamma^{\sigma \sigma'[11;\,\mbox{\scriptsize I}]}_{j_1 j_2;\, j_3 j_4}
(\epsilon,  \epsilon'; \omega) \,\right]
\nonumber \\
& & - f(\epsilon') \, 
\Gamma^{\sigma \sigma'[11;\,\mbox{\scriptsize I}]}_{j_1 j_2;\, j_3 j_4}
(\epsilon,  \epsilon'; \omega) \;,
\label{eq:L11} 
\\
{\cal L}^{\sigma \sigma' \, [12]}_{j_1 j_2;\, j_3 j_4}
(\epsilon,  \epsilon'; \omega) 
&=& 
- \left[\,  f(\epsilon'+\omega)  - f(\epsilon') \,\right]
\,\Gamma^{\sigma \sigma'[12]}_{j_1 j_2;\, j_3 j_4}
(\epsilon,  \epsilon'; \omega) \;,
\label{eq:L12} 
\\
{\cal L}^{\sigma \sigma' \, [13]}_{j_1 j_2;\, j_3 j_4}
(\epsilon,  \epsilon'; \omega) 
&=& 
{\cal P} b(\epsilon'+\epsilon+\omega) 
\left[\, 
\Gamma^{\sigma \sigma'[13;\,\mbox{\scriptsize I}]}_{j_1 j_2;\, j_3 j_4}
(\epsilon,  \epsilon'; \omega) -
\Gamma^{\sigma \sigma'[13;\,\mbox{\scriptsize II}]}_{j_1 j_2;\, j_3 j_4}
(\epsilon,  \epsilon'; \omega) \,\right]
\nonumber \\
& &  + f(\epsilon'+\omega) \, 
\Gamma^{\sigma \sigma'[13;\,\mbox{\scriptsize I}]}_{j_1 j_2;\, j_3 j_4}
(\epsilon,  \epsilon'; \omega) \;,
\label{eq:L13} 
\\
{\cal L}^{\sigma \sigma' \, [21]}_{j_1 j_2;\, j_3 j_4}
(\epsilon,  \epsilon'; \omega) 
&=& 
- f(\epsilon') \,
\Gamma^{\sigma \sigma'[21]}_{j_1 j_2;\, j_3 j_4}
(\epsilon,  \epsilon'; \omega) \;,
\label{eq:L21} 
\\
{\cal L}^{\sigma \sigma' \, [22]}_{j_1 j_2;\, j_3 j_4}
(\epsilon,  \epsilon'; \omega) 
&=& 
f(\epsilon') \,
\Gamma^{\sigma \sigma'[22;\,\mbox{\scriptsize II}]}_{j_1 j_2;\, j_3 j_4}
(\epsilon,  \epsilon'; \omega) 
-f(\epsilon'+\omega) \,
\Gamma^{\sigma \sigma'[22;\,\mbox{\scriptsize IV}]}_{j_1 j_2;\, j_3 j_4}
(\epsilon,  \epsilon'; \omega) 
\nonumber \\ 
& & + {\cal P} b(\epsilon'-\epsilon)\, 
\left[\, 
\Gamma^{\sigma \sigma'[22;\,\mbox{\scriptsize II}]}_{j_1 j_2;\, j_3 j_4}
(\epsilon,  \epsilon'; \omega) -
\Gamma^{\sigma \sigma'[22;\,\mbox{\scriptsize III}]}_{j_1 j_2;\, j_3 j_4}
(\epsilon,  \epsilon'; \omega) \,\right]
\nonumber \\
& & + {\cal P} b(\epsilon'+\epsilon+\omega)\, 
\left[\, 
\Gamma^{\sigma \sigma'[22;\,\mbox{\scriptsize III}]}_{j_1 j_2;\, j_3 j_4}
(\epsilon,  \epsilon'; \omega) -
\Gamma^{\sigma \sigma'[22;\,\mbox{\scriptsize IV}]}_{j_1 j_2;\, j_3 j_4}
(\epsilon,  \epsilon'; \omega) \,\right] ,
\label{eq:L22} 
\\
{\cal L}^{\sigma \sigma' \, [23]}_{j_1 j_2;\, j_3 j_4}
(\epsilon,  \epsilon'; \omega) 
&=& 
 f(\epsilon'+\omega) \,
\Gamma^{\sigma \sigma'[23]}_{j_1 j_2;\, j_3 j_4}
(\epsilon,  \epsilon'; \omega) \;,
\label{eq:L23} 
\\
{\cal L}^{\sigma \sigma' \, [31]}_{j_1 j_2;\, j_3 j_4}
(\epsilon,  \epsilon'; \omega) 
&=& 
{\cal P} b(\epsilon'+\epsilon+\omega)
\left[\, 
\Gamma^{\sigma \sigma'[31;\,\mbox{\scriptsize II}]}_{j_1 j_2;\, j_3 j_4}
(\epsilon,  \epsilon'; \omega) -
\Gamma^{\sigma \sigma'[31;\,\mbox{\scriptsize I}]}_{j_1 j_2;\, j_3 j_4}
(\epsilon,  \epsilon'; \omega) \,\right]
\nonumber \\
& &  - f(\epsilon') \, 
\Gamma^{\sigma \sigma'[31;\,\mbox{\scriptsize I}]}_{j_1 j_2;\, j_3 j_4}
(\epsilon,  \epsilon'; \omega) \;,
\label{eq:L31} 
\\
{\cal L}^{\sigma \sigma' \, [32]}_{j_1 j_2;\, j_3 j_4}
(\epsilon,  \epsilon'; \omega) 
&=& 
- \left[\,  f(\epsilon'+\omega)  - f(\epsilon') \,\right]
\,\Gamma^{\sigma \sigma'[32]}_{j_1 j_2;\, j_3 j_4}
(\epsilon,  \epsilon'; \omega) \;,
\label{eq:L32} 
\\
{\cal L}^{\sigma \sigma' \, [33]}_{j_1 j_2;\, j_3 j_4}
(\epsilon,  \epsilon'; \omega) 
&=& 
{\cal P} b(\epsilon'-\epsilon) 
\left[\, 
\Gamma^{\sigma \sigma'[33;\,\mbox{\scriptsize I}]}_{j_1 j_2;\, j_3 j_4}
(\epsilon,  \epsilon'; \omega) -
\Gamma^{\sigma \sigma'[33;\,\mbox{\scriptsize II}]}_{j_1 j_2;\, j_3 j_4}
(\epsilon,  \epsilon'; \omega) \,\right]
\nonumber \\
& & + f(\epsilon'+\omega) \, 
\Gamma^{\sigma \sigma'[33;\,\mbox{\scriptsize I}]}_{j_1 j_2;\, j_3 j_4}
(\epsilon,  \epsilon'; \omega) \;.
\label{eq:L33} 
\end{eqnarray}
Here ${\cal P}$ denotes the Cauchy principal value,  
and $b(\epsilon) = [{\rm e}^{\beta \epsilon}-1]^{-1}$. 
The Fermi and Bose distribution functions 
came through the summation over ${\rm i}\varepsilon '$. 
We can now examine the behavior 
of $P_{j_4j_1}^{[k]}(\epsilon, \epsilon+\omega)$ 
for small $\omega$.
The first and third terms of the right-hand side 
of eq.\ (\ref{eq:P_ret1}) 
vanish at $\omega = 0$ because 
$\lambda_{R}^{[1]}(\epsilon, \epsilon) =0$ and 
$\lambda_{R}^{[3]}(\epsilon, \epsilon) =0$ as already mentioned.
Furthermore, due to 
the factor $\left[\,  f(\epsilon'+\omega)  - f(\epsilon') \,\right]$ 
appearing in eqs.\ (\ref{eq:L12}) and (\ref{eq:L32}), 
${\cal L}^{\sigma \sigma' \, [12]}_{j_1 j_2;\, j_3 j_4}
(\epsilon,  \epsilon'; 0) =0$ and 
${\cal L}^{\sigma \sigma' \, [32]}_{j_1 j_2;\, j_3 j_4}
(\epsilon,  \epsilon'; 0) =0$.
Therefore in the limit of $\omega \to 0$, 
$P_{j_4j_1}^{[1]}(\epsilon, \epsilon)=0$, 
$P_{j_4j_1}^{[3]}(\epsilon, \epsilon)=0$, and 
\begin{eqnarray}
P_{j_4j_1}^{[2]}(\epsilon, \epsilon) 
\,= \,  
 2 \sum_{\sigma'} \sum_{ j_2 j_3 \in C}
\int_{-\infty}^{\infty} {{\rm d}\epsilon' \over 2\pi {\rm i}}
\ {\cal L}^{\sigma \sigma' \, [22]}_{j_1 j_2;\, j_3 j_4}
(\epsilon,  \epsilon'; 0) 
\ G_{j_3,N}^-(\epsilon') 
\, 
\Gamma_R (\epsilon')
\, G_{N,j_2}^+(\epsilon' )
\;.
\label{eq:P_ret2} 
\end{eqnarray}
Similarly, through eq.\ (\ref{eq:Lambda_R}) we have,  
$\Lambda_{R;j_4j_1}^{[1]}(\epsilon, \epsilon)=0$, 
$\Lambda_{R;j_4j_1}^{[3]}(\epsilon, \epsilon)=0$, and  
$\Lambda_{R;j_4j_1}^{[2]}(\epsilon, \epsilon)
 = 2\Gamma_R(\epsilon)\, \delta_{N j_4}  \delta_{N j_1}
  + P_{j_4j_1}^{[2]}(\epsilon, \epsilon)$.
Therefore, 
the $\omega$-linear part of $K_{RL}(\omega + i 0^+ )$ 
comes only from the second term of the right-hand side 
of eq.\ (\ref{eq:K_ret}), and thus 
the conductance can be expressed as  
\begin{eqnarray}
g \  &=& \ {2 e^2 \over h} \, 
       \int_{-\infty}^{\infty} {\rm d}\epsilon 
       \left( -\,{\partial f \over \partial \epsilon} \right)
        {\cal T}(\epsilon)
      \;,
            \label{eq:cond1}
           \\
{\cal T}(\epsilon) &=& \ 
\sum_{ j_1 j_4\in C}
    2 \, \Gamma_L(\epsilon) \, 
    G_{1 j_4}^{-}(\epsilon) \, 
      \Lambda_{R;j_4j_1}^{[2]}(\epsilon, \epsilon) 
    \, G_{j_1 1}^{+}(\epsilon) \;.
\label{eq:T_eff_sp}
\end{eqnarray}
Among nine functions 
of ${\cal L}^{\sigma \sigma' \, [kk']}_{j_1 j_2;\, j_3 j_4}$
given by eqs.\ (\ref{eq:L11})--(\ref{eq:L33}), 
only one 
function ${\cal L}^{\sigma \sigma' \, [22]}_{j_1 j_2;\, j_3 j_4}$ is 
relevant to the conductance through eq.\ (\ref{eq:P_ret2}). 
We note that 
$\Lambda_{R;jj'}^{[2]}(\epsilon, \epsilon) 
= \left\{ \Lambda_{R;j'j}^{[2]}(\epsilon, \epsilon) \right\}^*$
due to the time-reversal symmetry (see also Appendix), 
and thus ${\cal T}(\epsilon)$ is real. 
The contributions of the bubble and vertex diagrams can be separated as 
${\cal T}(\epsilon) =
         {\cal T}^{(a)}(\epsilon) + {\cal T}^{(b)}(\epsilon)$,
\begin{eqnarray}
 {\cal T}^{(a)}(\epsilon) &=&
           \ 4\, \Gamma_L(\epsilon) 
           \, G_{1 N}^{-}(\epsilon)  
           \, \Gamma_R(\epsilon) 
           \, G_{N 1}^{+}(\epsilon) 
      \;,
      \label{eq:T_eff_a}
\\
{\cal T}^{(b)}(\epsilon) &=&
           \sum_{j_1 j_4 \in C}
      2
      \, \Gamma_L(\epsilon) \,  
      G_{1 j_4}^{-}(\epsilon)\, 
      P_{j_4j_1}^{[2]}(\epsilon, \epsilon) 
      \, G_{j_1 1}^{+}(\epsilon)  
\label{eq:T_eff_b}
\;.
\end{eqnarray}
At low temperatures, the conductance is determined 
by the behavior of ${\cal T}(\epsilon)$ at small $\epsilon$ 
because of the Fermi factor $-\, \partial f /\partial \epsilon$ 
in  eq.\ (\ref{eq:cond1}).
Specifically, 
the vertex contribution vanishes at zero temperature.\cite{ao7} 
This can be reexamined as follows:
at $\omega=0$, $\epsilon=0$ and $T=0$,
 the [2,2] vertex function vanishes, i.e.,  
 ${\cal L}^{\sigma \sigma' \, [22]}_{j_1 j_2;\, j_3 j_4} 
 (0,  \epsilon'; 0) =0$,  due to the restriction of the phase space 
by the distributions $f(\epsilon')$ and $b(\epsilon')$
in eq.\ (\ref{eq:L22}). 
Thus from eqs.\ (\ref{eq:P_ret2}) and (\ref{eq:T_eff_b}) we have, 
$P_{j_4j_1}^{[2]}(0, 0)=0$ and ${\cal T}^{(b)}(0)=0$, at zero temperature.

In this section, we have assumed 
the time-reversal symmetry of $U_{j_4 j_3; j_2 j_1}$,
but not assumed the precise form of this interaction. 
Thus, the present formulation can be applicable 
to the interaction which vanishes smoothly at the interfaces.  
We have also assumed that the two leads are noninteracting.   
Due to this assumption,
we have no wavefunction renormalization in eq.\ (\ref{eq:T_eff_sp}). 
If the interaction is switched on also in the leads, 
the wavefunction renormalization may be necessary 
to define the initial and final scattering states.\cite{Eliashberg}

\section{Current Conservation and Ward Identity}
\label{sec:WardIdentity}

In this section, we provide an alternative expression of the 
transmission probability ${\cal T}(\epsilon)$ 
in terms of a three-point correlation function, i.e., eq.\ (\ref{eq:T_eff2}).
This expression may be suitable for  
nonperturbative calculations such as numerical 
renormalization group\cite{Izumida3,Costi,Gerland} 
and quantum Monte Carlo method.\cite{Sakai}
The Lehmann representation of ${\cal T}(\epsilon)$ can be obtained
using the correlation function, and it is given in Appendix.  
We also discuss the current conservation in terms of 
the generalized Ward identity eq.\ (\ref{eq:Ward1}).
This identity can be rewritten in the form of eq.\ (\ref{eq:Ward3_ret_0_b}), 
and it shows a relationship between the self-energy and current vertex.

The three-point correlation functions of 
the charge and currents are defined by
\begin{eqnarray} 
\Phi_{C;jj'}(\tau; \tau_1, \tau_2) 
&=&    \left \langle  T_{\tau} \, \delta\rho_C^{\phantom{0}}(\tau)\,
 c^{\phantom{\dagger}}_{j \sigma} (\tau_1) \, 
 c^{\dagger}_{j' \sigma} (\tau_2)                      
 \right \rangle ,
\label{eq:Phi_C}
\\
\Phi_{L;jj'}(\tau; \tau_1, \tau_2) 
&=&    \left \langle T_{\tau} \, J_L(\tau)\,
 c^{\phantom{\dagger}}_{j \sigma} (\tau_1) \, 
 c^{\dagger}_{j' \sigma} (\tau_2)                      
 \right \rangle ,
\label{eq:Phi_L}
\\
\Phi_{R;jj'}(\tau; \tau_1, \tau_2) 
&=&    \left \langle  T_{\tau} \, J_R(\tau)\,
 c^{\phantom{\dagger}}_{j \sigma} (\tau_1) \, 
 c^{\dagger}_{j' \sigma} (\tau_2) 
     \right \rangle ,
\label{eq:Phi_R}
\end{eqnarray} 
where 
$\delta \rho_C^{\phantom{0}} \equiv \rho_C^{\phantom{0}} 
 - \langle \rho_C^{\phantom{0}} \rangle$.
The Fourier transform of these functions are given by
\begin{eqnarray}
\Phi_{\gamma:jj'}(\tau; \tau_1, \tau_2) 
= {1 \over \beta^2} 
\sum_{{\rm i}\varepsilon , {\rm i}\nu} 
\Phi_{\gamma:jj'}
({\rm i}\varepsilon , {\rm i}\varepsilon  + {\rm i}\nu)
\, {\rm e}^{-{\rm i}\,\varepsilon (\tau_1 - \tau)} 
\, {\rm e}^{-{\rm i}\,(\varepsilon + \nu) (\tau - \tau_2)} 
\;,
\end{eqnarray}
for $\gamma = C, L, R$. 
In what follows, we will discuss the correlation functions 
in the central region assuming $jj' \in C$. 
The right-hand side of eqs.\ (\ref{eq:Phi_L}) and (\ref{eq:Phi_R}) 
contain the operators with respect to sites in the reservoirs, 
$0$ and $N+1$, through $J_L$ and $J_R$.  
The contributions of these two sites 
can be rewritten in terms of the Green's functions with respect 
to the adjacent sites  $1$ and $N$  
using eq.\ (\ref{eq:rec_lead}) with the Feynman diagrams 
for $\Phi_{\gamma:jj'}({\rm i}\varepsilon, {\rm i}\varepsilon  + {\rm i}\nu)$. 
Therefore, these correlation functions 
can be expressed in terms of the current 
vertex introduced in \S \ref{sec:CONDUCTANCE} as
\begin{eqnarray}
  \Phi_{\gamma;jj'}({\rm i}\varepsilon , {\rm i}\varepsilon  + {\rm i}\nu)
  = \sum_{j_4j_1 \in C}
  G_{jj_4}({\rm i}\varepsilon )\, 
\Lambda_{\gamma;j_4j_1}({\rm i}\varepsilon , {\rm i}\varepsilon  + {\rm i}\nu)
  \, G_{j_1j'}({\rm i}\varepsilon  + {\rm i}\nu) 
  \;.
  \label{eq:Lambda}
\end{eqnarray}
Here $\Lambda_{R;j_4j_1}$ is the current vertex for $J_R$ given by
eqs.\ (\ref{eq:Lambda_R}) and (\ref{eq:P_M}). 
The functions $\Lambda_{L;j_4j_1}$ and $\Lambda_{C;j_4j_1}$  
are defined by the similar expressions;    
the bare vertices are given by 
 $\lambda_L({\rm i}\varepsilon, {\rm i}\varepsilon  + {\rm i}\nu) 
\, \delta_{1j_4} \, \delta_{1j_1}$ and $\delta_{j_4j_1}$, respectively. 
Thus, the transmission probability, eq.\ (\ref{eq:T_eff_sp}), 
corresponds to the analytic continuation of eq.\ (\ref{eq:Lambda}) 
in the region, 
${\rm i}\varepsilon  + {\rm i}\nu \to \epsilon +\omega +{\rm i}0^+$  
and 
${\rm i}\varepsilon  \to \epsilon -i 0^+$, as 
\begin{eqnarray}
{\cal T}(\epsilon)  &=&     
  2\, \Gamma_L(\epsilon) \, \Phi_{R;11}^{[2]}(\epsilon, \epsilon) 
\;.
\label{eq:T_eff2}
\end{eqnarray}
The Lehmann representation 
of $\Phi_{R;11}^{[2]}(\epsilon, \epsilon)$ is presented in Appendix.

In the rest of this section, 
we discuss the current conservation in terms of these correlation 
functions. 
Using the equation of continuity in the Matsubara form 
$- \left({\partial /\partial \tau}\right) \delta \rho_C^{\phantom{0}}
+ {\rm i}\, J_R - {\rm i}\,J_L = 0$, the generalized Ward identity 
 is obtained as\cite{Schrieffer} 
\begin{eqnarray}
& & -\, {\partial \over  \partial \tau }\,  
\Phi_{C;jj'}(\tau; \tau_1, \tau_2) 
\,
+ {\rm i}\,  
\Phi_{R;jj'}(\tau; \tau_1, \tau_2) 
- {\rm i}\,
\Phi_{L;jj'}(\tau; \tau_1, \tau_2) 
\nonumber \\ 
&=& \  \, \delta (\tau - \tau_2) \, G_{jj'}(\tau_1,\tau) 
        -\, \delta (\tau_1 - \tau) \, G_{jj'}(\tau,\tau_2) 
\;.
\label{eq:Ward1} 
\end{eqnarray}
The Fourier transform of this identity can be expressed, 
using  $N\times N$ matrices, as  
\begin{eqnarray}
 {\rm i}\nu \, 
 \mbox{\boldmath $\Phi$}_C
 ({\rm i}\varepsilon , {\rm i}\varepsilon  + {\rm i}\nu) + {\rm i}\,  
\mbox{\boldmath $\Phi$}_R
({\rm i}\varepsilon, {\rm i}\varepsilon + {\rm i}\nu)
 - {\rm i}\, 
\mbox{\boldmath $\Phi$}_L
({\rm i}\varepsilon, {\rm i}\varepsilon + {\rm i}\nu)
\, = \,
 \mbox{\boldmath $G$}({\rm i}\varepsilon ) 
 -
 \mbox{\boldmath $G$}({\rm i}\varepsilon +{\rm i}\nu) 
 .
 \label{eq:Ward2} 
\end{eqnarray}
Here $\mbox{\boldmath $G$}(z) = \{G_{jj'}(z)\}$, 
$\mbox{\boldmath $\Phi$}_{\gamma}(z,z+w)  = \{\Phi_{\gamma;jj'}(z,z+w)\}$,
and 
$\mbox{\boldmath $\Lambda$}_{\gamma}(z,z+w) = \{\Lambda_{\gamma;jj'}(z,z+w)\}$
for $jj' \in  C$.
In this notation, eq.\ (\ref{eq:Lambda}) is written as  
$
 \mbox{\boldmath $\Phi$}_{\gamma}(z, z + w)
=  \mbox{\boldmath $G$}(z)\, 
\mbox{\boldmath $\Lambda$}_{\gamma}(z, z + w)
\, \mbox{\boldmath $G$}(z + w) 
$. Therefore eq.\ (\ref{eq:Ward2}) can be rewritten as
\begin{eqnarray}
& & 
{\rm i}\nu \, 
\mbox{\boldmath $\Lambda$}_C
({\rm i}\varepsilon , {\rm i}\varepsilon  + {\rm i}\nu)
+ {\rm i}\, 
\mbox{\boldmath $\Lambda$}_R
({\rm i}\varepsilon, {\rm i}\varepsilon + {\rm i}\nu)
- {\rm i}\, \mbox{\boldmath $\Lambda$}_L
({\rm i}\varepsilon, {\rm i}\varepsilon + {\rm i}\nu)
=
\left\{\mbox{\boldmath $G$}({\rm i}\varepsilon +{\rm i}\nu)\right\}^{-1}  
-
\left\{\mbox{\boldmath $G$}({\rm i}\varepsilon )\right\}^{-1} .
\label{eq:Ward3} 
\end{eqnarray}
Furthermore, the Dyson equation eq.\ (\ref{eq:Dyson}) can be expressed as 
$\left\{\mbox{\boldmath $G$}(z)\right\}^{-1}  
= 
 z \, \mbox{\boldmath $1$} 
 -  \mbox{\boldmath ${\cal H}$}_C^0  
- \mbox{\boldmath ${\cal V}$}_{mix}(z)  
- \mbox{\boldmath $\Sigma$}(z) $:    
\begin{eqnarray}
 \mbox{\boldmath ${\cal H}$}_C^0 \ &=& \  
\left [ \,
 \matrix { -t_{11}^C-\mu    & -t_{12}^C      & \cdots  &               \cr
            -t_{21}^C        & -t_{22}^C -\mu &         &               \cr
            \vdots          &               & \ddots  &               \cr 
                            &               &         & -t_{NN}^C -\mu \cr
          }
          \, \right ]  
\;,
\label{eq:matrix_H0}
\\
\mbox{\boldmath ${\cal V}$}_{mix}(z) \  &=& \ 
 \left [ \, \matrix { 
 v_L^{2} \, \mbox{\sl g}_L(z) & 0 & \cdots &0 & 0 \cr
 0           &    0     & \cdots   &  0 & 0     \cr
 \vdots      &  \vdots  & \ddots   &  \vdots & \vdots  \cr 
 0 & 0 & \cdots & 0 & 0 \cr
 0 & 0 & \cdots & 0 & v_R^{2} \, \mbox{\sl g}_R(z) \cr
                    }
 \, \right ]  
\label{eq:V_mix} 
\;,
\end{eqnarray}
and $\mbox{\boldmath $\Sigma$}(z) = \{\Sigma_{jj'}(z)\}$. 
The mixing matrix $\mbox{\boldmath ${\cal V}$}_{mix}$ has only two
non-zero elements, because among the sites in the central region 
only the two sites at $1$ and $N$ are connected to the reservoirs.
Note that eq.\ (\ref{eq:Ward3}) shows   
a relationship among the self-energy and current vertices,
which has to be satisfied in approximate calculations 
in order to obtain current-conserving results.
We now consider the analytic continuation of eq.\ (\ref{eq:Ward3}) 
in the region 
${\rm i}\varepsilon  + {\rm i}\nu \to \epsilon +\omega +{\rm i}0^+$ and 
${\rm i}\varepsilon  \to \epsilon -i 0^+$.
In the limit of $\omega \to 0$, it is given by
$
\mbox{\boldmath $\Lambda$}_R^{[2]}
(\epsilon,  \epsilon)
- \mbox{\boldmath $\Lambda$}_L^{[2]}
(\epsilon,  \epsilon) = 
-2\, \mbox{Im}\, \mbox{\boldmath ${\cal V}$}_{mix}^+(\epsilon)
-2\, \mbox{Im}\, \mbox{\boldmath $\Sigma$}^+(\epsilon) 
$.
This relation can be rewritten further, 
using eq.\ (\ref{eq:P_ret2}) and the corresponding expression 
for $\Lambda_{L;j_4j_1}^{[2]}(\epsilon,  \epsilon)$, as 
\begin{eqnarray}
& & \!\!\!\!\!\!\!\!\!\!\!\!\!
-\, \mbox{Im}\, \Sigma^+_{j_4 j_1}(\epsilon) \ = 
\nonumber \\
& & \sum_{\sigma'} \sum_{ j_2j_3 \in C}
\int_{-\infty}^{\infty} {{\rm d}\epsilon' \over 2\pi {\rm i}}
\, {\cal L}^{\sigma \sigma' \, [22]}_{j_1 j_2;\, j_3 j_4}
(\epsilon,  \epsilon'; 0) 
\left[\,
G_{j_3 N}^-(\epsilon') \, 
\Gamma_R (\epsilon')\, 
G_{N j_2}^+(\epsilon' ) 
+
G_{j_3 1}^-(\epsilon') 
\, \Gamma_L (\epsilon') 
\, G_{1 j_2}^+(\epsilon' ) 
\,\right] .
\label{eq:Ward3_ret_0_b} 
\end{eqnarray}
We note that 
eq.\ (\ref{eq:Ward3_ret_0_b}) can be regarded 
as a relation between the quasi-particles damping
and transport relaxation time.
At $T=0$, the damping rate due to the inelastic scattering vanishes 
 $\mbox{Im}\, \Sigma^+_{j_4 j_1}(0)=0$,
and it is connecting to the property of the vertex part
${\cal L}^{\sigma \sigma' \, [22]}_{j_1 j_2;\, j_3 j_4}
(0,  \epsilon'; 0) =0$.

Specifically, in the single impurity case $N=1$, 
eq.\ (\ref{eq:Ward2}) is a scalar equation with respect 
to the impurity site. 
It is written,
performing the analytic continuation, as  
$
\Phi_R^{[2]}(\epsilon,  \epsilon)
- \Phi_L^{[2]}(\epsilon,  \epsilon) = 
G^-(\epsilon) - G^+(\epsilon) 
$. 
Furthermore, if the mixing terms satisfy the condition 
$\Gamma_L(\epsilon) = \lambda \,\Gamma_R(\epsilon)$,
we have $\Phi_L^{[2]}(\epsilon,  \epsilon) = 
- \lambda\, \Phi_R^{[2]}(\epsilon, \epsilon)$. Then 
the conductance can be written in a well-known expression;\cite{MW,HDW2}
\begin{equation}
g_{\rm single}^{\phantom{0}} \,  
 =  {2 e^2 \over h} \, 
\int_{-\infty}^{\infty} 
  {\rm d}\epsilon  
\, \left(- {\partial f \over \partial \epsilon} \right)
 \frac{4\, \Gamma_L \Gamma_R}{\Gamma_R + \Gamma_L} 
 \left[
  - \mbox{Im}\,G^+(\epsilon) \right ]  .
\label{eq:cond_single}
\end{equation}

\section{Application to the Anderson Impurities}
\label{sec:2nd_order}

\begin{figure}
\rule{4.2cm}{0cm}
\centerline{ \vbox{ \epsfxsize=80mm \epsfbox {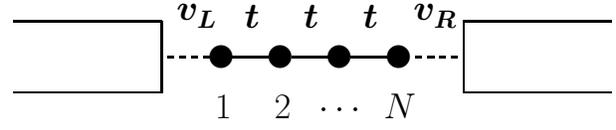}}
}
\caption{ Schematic picture of the model. 
} 
\label{fig:lattice}
\end{figure}

In this section we apply the formulation 
described in previous sections 
to a linear chain of the Anderson impurities. 
A schematic picture of this model is shown in Fig.\ \ref{fig:lattice}: 
the system consists of a series of $N$ impurities 
at the center and two reservoirs. 
This system may be considered as a model for a series 
of quantum dots or atomic wires of nanometer size.
For $N=1$ and  $N=2$, the conductance has already been 
calculated with efficient numerical methods 
such as the numerical renormalization group 
\cite{Izumida3} and quantum Monte Carlo methods.\cite{ao6,Sakai} 
Our interest here is how 
the transmission probability ${\cal T}(\epsilon)$ depends on 
the frequency and temperature.
In the previous work we have calculated the conductance 
using the order $U^2$ self-energy at $T=0$, 
and found that the off-diagonal (non-local) elements of the self-energy 
play an important role on the transport properties.\cite{ao7,ao9}
In order to calculate the conductance at finite temperatures, 
the contribution of the vertex part eq.\ (\ref{eq:T_eff_b})
has to be taking into account carefully. 
In the present work, we calculate all the matrix elements of the order $U^2$ 
self-energy and current vertex 
which satisfy the generalized Ward identity eq.\ (\ref{eq:Ward3_ret_0_b}) 
to obtain ${\cal T}(\epsilon)$ for $N=1,2, 3$, and $4$.
We note that the second-order perturbation theory has been 
applied to transport through 
the single Anderson impurity\cite{HDW2,YMF,MiiMakoshi,TakagiSaso}
and related systems\cite{PFA,KKNO} in somewhat different contexts.

In this section 
the parameters of the Hamiltonian defined by eq.\ (\ref{eq:H}) are 
given explicitly as follows.  
For $t_{ij}^C$, we take the nearest-neighbor elements 
to be $t$ and all other off-diagonal elements to be zero.
We assume $U_{j_4 j_3; j_2 j_1}$ to be an onsite repulsion $U$. 
Furthermore, we will concentrate on the electron-hole symmetric case
taking the parameters to be $\mu =0$ and $\epsilon_0 + U/2 =0$, 
where $\epsilon_0$ is the onsite energy $-t_{ii}^C=\epsilon_0$.
Then the Dyson equation can be written in the $N\times N$ matrix form 
as,  $\left\{ \mbox{\boldmath $G$}(z)\right\}^{-1}  
=    \left\{ \mbox{\boldmath $G^0$}(z)\right\}^{-1}   
 -  \mbox{\boldmath $\Sigma$}(z)$, 
\begin{eqnarray}
\left\{ 
\mbox{\boldmath $G^0$}(\epsilon+{\rm i}0^+)
\right\}^{-1}  
&=& 
\left [ \,
 \matrix {\epsilon - v_L^2\, \mbox{\sl g}_{L}^{+}(\epsilon) 
           & t & & & \mbox{\Large \bf 0}  \cr
            t &\epsilon & t  &  &  \cr
               & \ddots \qquad & \ddots & \qquad \ddots &  \cr 
               &  & t &\epsilon &t \rule{1.0cm}{0cm} \cr
          \mbox{\Large \bf 0} & & & t & 
          \epsilon - v_R^2\, \mbox{\sl g}_{R}^{+}(\epsilon)  \cr
          }
          \, \right ]  .
\label{eq:G_0_matrix}
\end{eqnarray}
Here 
$v_{\alpha}^2 \, \mbox{\sl g}_{\alpha}^{+}(\epsilon)$ 
for $\alpha=L,R$ corresponds to the mixing term 
eq.\ (\ref{eq:V_mix}) due to the coupling with the reservoirs, 
and the imaginary part   
$\Gamma_{\alpha}(\epsilon) = -\, \mbox{Im} 
\left[ v_{\alpha}^2 \, \mbox{\sl g}_{\alpha}^{+}(\epsilon) \right]$ is 
finite because each reservoirs has a continuous energy spectrum.  
Thus,
the unperturbed Green's function $\mbox{\boldmath $G^0$}(z)$  
describes a system of $N$ resonant states which have finite level width. 
Note that the Hartree-Fock term of $U$ 
is already included in eq.\ (\ref{eq:G_0_matrix}), 
and thus $\mbox{\boldmath $\Sigma$}(z)$ is the self-energy 
due to the interaction Hamiltonian,    
$
{\cal H}_C^{int} = U \sum_{i=1}^{N} \left[\, 
                 n_{i \uparrow}\, n_{i \downarrow}
-   (   n_{i \uparrow}
                   + n_{i \downarrow} )/2
            \,\right]
$, where $n_{i \sigma} = 
 c^{\dagger}_{i \sigma} c^{\phantom{\dagger}}_{i \sigma}$.

\begin{figure}
\vspace{-1cm}
\centerline{ \vbox{ \epsfxsize=180mm \epsfbox {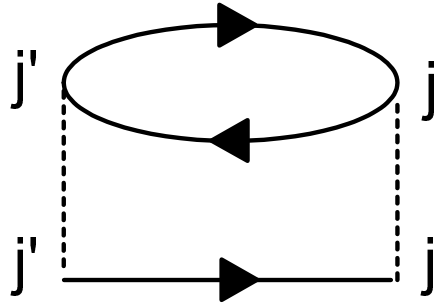}}
}
\vspace{-19.5cm}
\caption{The order $U^2$ self-energy $\Sigma_{jj'}$.}
\label{fig:diagramSelf}
\end{figure}

\begin{figure}
\vspace{-2cm}
\centerline{ \vbox{ \epsfxsize=180mm \epsfbox {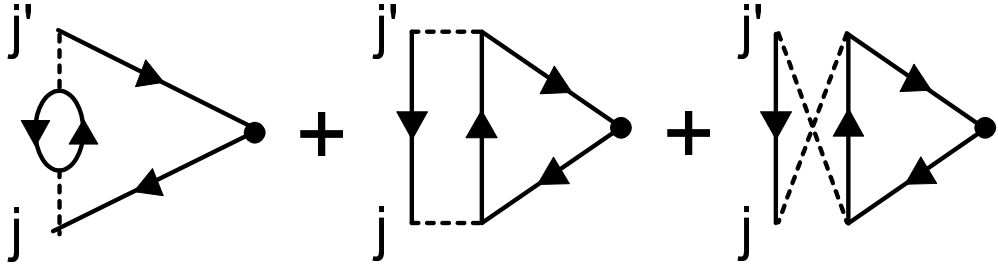}}
}
\vspace{-19cm}
\caption{The order $U^2$ current vertex $P_{jj'}$.}
\label{fig:diagramVertex}
\end{figure}

We now calculate the self-energy and current vertex   
in the second-order perturbation
with respect to ${\cal H}_C^{int}$.
The order $U^2$ self-energy is shown in Fig.\ \ref{fig:diagramSelf}, 
and is given by 
\begin{eqnarray}
 \mbox{Im}\, \Sigma_{jj'}^+( \epsilon) 
    &=& \left({U \over \pi}\right)^2 
         \int_{-\infty}^{\infty}
         {\rm d}\epsilon_1 
         \int_{-\infty}^{\infty} 
          {\rm d}\epsilon_2
       \  \mbox{Im}\,  G^{0+}_{jj'}(\epsilon_1)  
       \, \mbox{Im}\, G^{0+}_{j'j}(\epsilon_2) 
       \, \mbox{Im}\, G^{0+}_{jj'}(\epsilon_2 - \epsilon_1 + \epsilon) 
\nonumber \\
& &  
\times \  \Bigl \{\,
f(\epsilon_2)\,\left[\,1-f(\epsilon_1)\, \right]\,
\left[\,1-f(\epsilon_2 - \epsilon_1 + \epsilon) \,\right] 
+ \left[\,1-f(\epsilon_2) \,\right] \, f(\epsilon_1)
 \, f(\epsilon_2 - \epsilon_1 + \epsilon)  
\,\Bigr \},
  \rule{0.5cm}{0cm}
 \label{eq:Self2_Im}
 \\ 
\mbox{Re}\, \Sigma_{jj'}^+( \epsilon) 
    &=& \left({U \over \pi}\right)^2 
         \int_{-\infty}^{\infty}
         {\rm d}\epsilon_1 
         \int_{-\infty}^{\infty} 
          {\rm d}\epsilon_2 \, 
\nonumber \\
& & 
 \Bigl \{\,
  \mbox{Re}\,  G^{0+}_{jj'}(\epsilon_1)  
       \, \mbox{Im}\, G^{0+}_{j'j}(\epsilon_2) 
       \, \mbox{Im}\, G^{0+}_{jj'}(\epsilon_2 - \epsilon_1 + \epsilon) 
\, \left[\,1-f(\epsilon_2 - \epsilon_1 + \epsilon) \,\right] \, f(\epsilon_2) 
\nonumber \\
& & \, - \, \mbox{Im}\,  G^{0+}_{jj'}(\epsilon_1)  
       \, \mbox{Im}\, G^{0+}_{j'j}(\epsilon_2) 
       \, \mbox{Re}\, G^{0+}_{jj'}(\epsilon_2 - \epsilon_1 + \epsilon) 
\, f(\epsilon_1)\, f(\epsilon_2) 
\nonumber \\
& & \, - \,  
       \mbox{Im}\,  G^{0+}_{jj'}(\epsilon_1)  
       \, \mbox{Re}\, G^{0+}_{j'j}(\epsilon_2) 
       \, \mbox{Im}\, G^{0+}_{jj'}(\epsilon_2 - \epsilon_1 + \epsilon) 
\,  f(\epsilon_1) \,f(\epsilon_2 - \epsilon_1 + \epsilon) 
\, \Bigr \} . 
\label{eq:Self2_Re} 
\end{eqnarray}
The order $U^2$ current vertex 
 $P_{j_4 j_1}({\rm i}\varepsilon , {\rm i}\varepsilon  +{\rm i}\nu )$,  
which satisfies the generalized Ward identity eq.\ (\ref{eq:Ward3})
with this order $U^2$ self-energy, is shown 
in Fig.\ \ref{fig:diagramVertex}. 
The contribution of the four-point vertex part is given by 
\begin{eqnarray}
 \Gamma^{\sigma \sigma'}_{j_1 j_2;\, j_3 j_4}
  ({\rm i}\varepsilon , {\rm i}\varepsilon '; {\rm i}\nu) 
&=& \ \delta_{\sigma \sigma'}\, \delta_{j_1 j_2}\, \delta_{j_3 j_4}\, 
  U^2  X_{j_4 j_1}({\rm i}\varepsilon  - {\rm i}\varepsilon ') 
\nonumber \\
& & 
+ \, \delta_{\sigma -\sigma'}\, \delta_{j_1 j_2}\, \delta_{j_3 j_4}\, 
  U^2   X_{j_4 j_1}({\rm i}\varepsilon  - {\rm i}\varepsilon ') 
\nonumber \\
& & + \, \delta_{\sigma -\sigma'}\, \delta_{j_1 j_3}\, \delta_{j_2 j_4}\, 
  U^2 \, Y_{j_4 j_1}({\rm i}\varepsilon  + {\rm i}\varepsilon ' + {\rm i}\nu) 
  \;,
 \label{eq:2nd_vertex} 
\\
 X_{j_4 j_1}({\rm i}\nu) &=& -\,{1 \over \beta} 
\,\sum_{{\rm i}\varepsilon ''}
\, G^0_{j_4 j_1}({\rm i}\nu + {\rm i}\varepsilon '' ) 
\, G^0_{j_1 j_4}({\rm i}\varepsilon '')  \;,
\label{eq:p-h}
\\
 Y_{j_4 j_1}({\rm i}\nu) &=& -\, {1 \over \beta} 
\,\sum_{{\rm i}\varepsilon ''}
\, G^0_{j_4 j_1}({\rm i}\nu - {\rm i}\varepsilon '') 
\, G^0_{j_4 j_1}({\rm i}\varepsilon '')  \;,
\label{eq:p-p}
\end{eqnarray}
where each terms in the right-hand side of 
eq.\ (\ref{eq:2nd_vertex}) 
corresponds to each diagrams in Fig.\ \ref{fig:diagramVertex}. 
The analytic continuation of 
 $P_{j_4 j_1}$ for ${\rm i}\varepsilon  
 + {\rm i}\nu \to \epsilon +\omega +{\rm i}0^+$ and   
${\rm i}\varepsilon  \to \epsilon -i 0^+$ can be performed explicitly
using this four-point vertex. 
That yields the expression corresponding to eq.\ (\ref{eq:P_ret2}) 
with 
\begin{eqnarray}
 {\cal L}^{\sigma \sigma' \, [22]}_{j_1 j_2;\, j_3 j_4}
  (\epsilon, \epsilon'; \omega) 
&=& \    
\left[\, \delta_{\sigma \sigma'} \, +\, \delta_{\sigma -\sigma'} \,\right]\, 
\delta_{j_1 j_2}\, \delta_{j_3 j_4}\, U^2 
\nonumber
\\ 
& & 
\times 
\, \Bigl\{ 
\,  f(\epsilon')\, X_{j_4 j_1}^-(\epsilon - \epsilon') 
-\,f(\epsilon' + \omega)\,X_{j_4 j_1}^+(\epsilon - \epsilon') 
\nonumber \\ 
& & \qquad 
 -\, b(\epsilon' - \epsilon) 
  \left[\,X_{j_4 j_1}^+(\epsilon - \epsilon') 
- X_{j_4 j_1}^-(\epsilon - \epsilon') \,\right]
\, \Bigr\}
\nonumber \\ 
& & + \,
\delta_{\sigma -\sigma'}\, \delta_{j_1 j_3}\, \delta_{j_2 j_4}\, U^2
\nonumber 
\\ 
& & \times 
\, \Bigl\{ 
\,  f(\epsilon')\, Y_{j_4 j_1}^+(\epsilon + \epsilon' + \omega) 
-\,f(\epsilon' + \omega)\,Y_{j_4 j_1}^-(\epsilon + \epsilon' + \omega) 
\nonumber \\ 
& & \qquad 
 +\, b(\epsilon + \epsilon' + \omega) 
\left[\,Y_{j_4 j_1}^+(\epsilon + \epsilon' + \omega) 
- Y_{j_4 j_1}^-(\epsilon + \epsilon' + \omega) \,\right]
\, \Bigr\}
.
\label{eq:2nd_vertex_22}
\end{eqnarray}
Here $X_{j_4 j_1}^{\pm}(\omega)$ and $Y_{j_4 j_1}^{\pm}(\omega)$ 
is the analytic continuation of 
eqs.\ (\ref{eq:p-h}) and (\ref{eq:p-p}), respectively, 
\begin{eqnarray}
\mbox{Im}\, X_{j_4 j_1}^+(\omega)
&=&  \int_{-\infty}^{\infty} {{\rm d}\epsilon \over \pi}
\ \left[\, f(\epsilon) - f(\epsilon+\omega) \,\right]
\,
\mbox{Im}\, G^{0+}_{j_4 j_1}(\epsilon + \omega)  
\,\mbox{Im}\, G^{0+}_{j_1 j_4}(\epsilon)  
\;,
\label{eq:p-h_Im}
\\
\mbox{Im}\, Y_{j_4 j_1}^+(\omega)
&=& \int_{-\infty}^{\infty} {{\rm d}\epsilon \over \pi}
\ \left[\, f(\epsilon) - f(\epsilon - \omega) \,\right]
\,
\mbox{Im}\, G^{0+}_{j_4 j_1}(\omega - \epsilon)  
\ \mbox{Im}\, G^{0+}_{j_4 j_1}(\epsilon)  
\label{eq:p-p_Im}
\;.
\end{eqnarray}
Using eqs.\ (\ref{eq:2nd_vertex_22})--(\ref{eq:p-p_Im}), 
the analytic continuation $P^{[2]}_{j_4 j_1}$ can be 
obtained up to order $U^2$ as  
\begin{eqnarray}
 P^{[2]}_{j_4 j_1}(\epsilon, \epsilon) 
    &=& 
    4 \left({U \over \pi}\right)^2 
         \int_{-\infty}^{\infty}
         {\rm d}\epsilon_1 
         \int_{-\infty}^{\infty} 
          {\rm d}\epsilon_2
       \  G^{0-}_{j_4 N }(\epsilon_1) 
       \,  G^{0+}_{N j_1}(\epsilon_1)  
       \, \Gamma_R(\epsilon_1)  
\nonumber  \\  
    & & \times \ \mbox{Im}\, G^{0+}_{j_1 j_4}(\epsilon_2) 
       \ \mbox{Im}\, G^{0+}_{j_4 j_1}(\epsilon_2 - \epsilon_1 + \epsilon) 
\nonumber \\
& &  
\times \  \Bigl \{\,
f(\epsilon_2)\,\left[\,1-f(\epsilon_1)\, \right]\,
\left[\,1-f(\epsilon_2 - \epsilon_1 + \epsilon) \,\right] 
+ \left[\,1-f(\epsilon_2) \,\right] \, f(\epsilon_1)
 \, f(\epsilon_2 - \epsilon_1 + \epsilon)  
\,\Bigr \}
\nonumber \\
 & & -2   \left({U \over \pi}\right)^2 
         \int_{-\infty}^{\infty}
         {\rm d}\epsilon_1 
         \int_{-\infty}^{\infty} 
          {\rm d}\epsilon_2
       \  G^{0-}_{j_1 N }(\epsilon_2)
       \,  G^{0+}_{N j_4}(\epsilon_2)  
       \, \Gamma_R(\epsilon_2)  
       \nonumber \\
& & \times  \ \mbox{Im}\, G^{0+}_{j_4 j_1}(\epsilon_1) 
       \ \mbox{Im}\, G^{0+}_{j_4 j_1}(\epsilon_2 - \epsilon_1 + \epsilon) 
\nonumber \\
& &  
\times \  \Bigl \{\,
f(\epsilon_2)\,\left[\,1-f(\epsilon_1)\, \right]\,
\left[\,1-f(\epsilon_2 - \epsilon_1 + \epsilon) \,\right] 
+ \left[\,1-f(\epsilon_2) \,\right] \, f(\epsilon_1)
 \, f(\epsilon_2 - \epsilon_1 + \epsilon)  
\,\Bigr \}
.
  \rule{0.7cm}{0cm}
 \label{eq:vertex2} 
\end{eqnarray}
We note that 
the first and second diagrams in Fig.\ \ref{fig:diagramVertex}
give the same contributions. 
Furthermore, 
the Green's functions have some special properties
in the electron-hole symmetric case: 
 $G_{ij}^{+}(-\epsilon) = (-1)^{i-j+1} \,G_{ij}^{-}(\epsilon)$, 
 $\Sigma_{ij}^{+}(-\epsilon) = (-1)^{i-j+1} \,\Sigma_{ij}^{-}(\epsilon)$, 
and $P^{[2]}_{ij}(-\epsilon, -\epsilon) 
= (-1)^{i-j}\, P^{[2]}_{ji}(\epsilon, \epsilon)$.
Then the contributions of the second and third diagrams 
cancel each other out, and eq.\ (\ref{eq:vertex2}) simplifies as
\begin{eqnarray}
 P^{[2]}_{j_4 j_1}(\epsilon, \epsilon) 
    &=& 
    2 \left( {U \over \pi} \right)^2 
         \int_{-\infty}^{\infty}
         {\rm d}\epsilon_1 
         \int_{-\infty}^{\infty} 
          {\rm d}\epsilon_2
       \  G^{0-}_{j_4 N }(\epsilon_1) 
       \,  G^{0+}_{N j_1}(\epsilon_1)  
       \, \Gamma_R(\epsilon_1)  
\nonumber  \\  
    & & \times \ \mbox{Im}\, G^{0+}_{j_1 j_4}(\epsilon_2) 
       \ \mbox{Im}\, G^{0+}_{j_4 j_1}(\epsilon_2 - \epsilon_1 + \epsilon) 
\nonumber \\
& &  
\times \  \Bigl \{\,
f(\epsilon_2)\,\left[\,1-f(\epsilon_1)\, \right]\,
\left[\,1-f(\epsilon_2 - \epsilon_1 + \epsilon) \,\right] 
+ \left[\,1-f(\epsilon_2) \,\right] \, f(\epsilon_1)
 \, f(\epsilon_2 - \epsilon_1 + \epsilon)  
\,\Bigr \}
.
  \rule{0.7cm}{0cm}
 \label{eq:vertex3} 
\end{eqnarray}
Note that the restriction of the integration region 
due to the Fermi functions has the same form 
with that in eq.\ (\ref{eq:Self2_Im}). Therefore,
$P^{[2]}_{j_4 j_1}(\epsilon, \epsilon)$ and 
 $\mbox{Im}\, \Sigma_{jj'}^+( \epsilon)$ 
show the similar $T$ and $\epsilon$ dependences.

\begin{figure}
\vspace{-2cm}
\centerline{ \vbox{ \epsfxsize=160mm \epsfbox {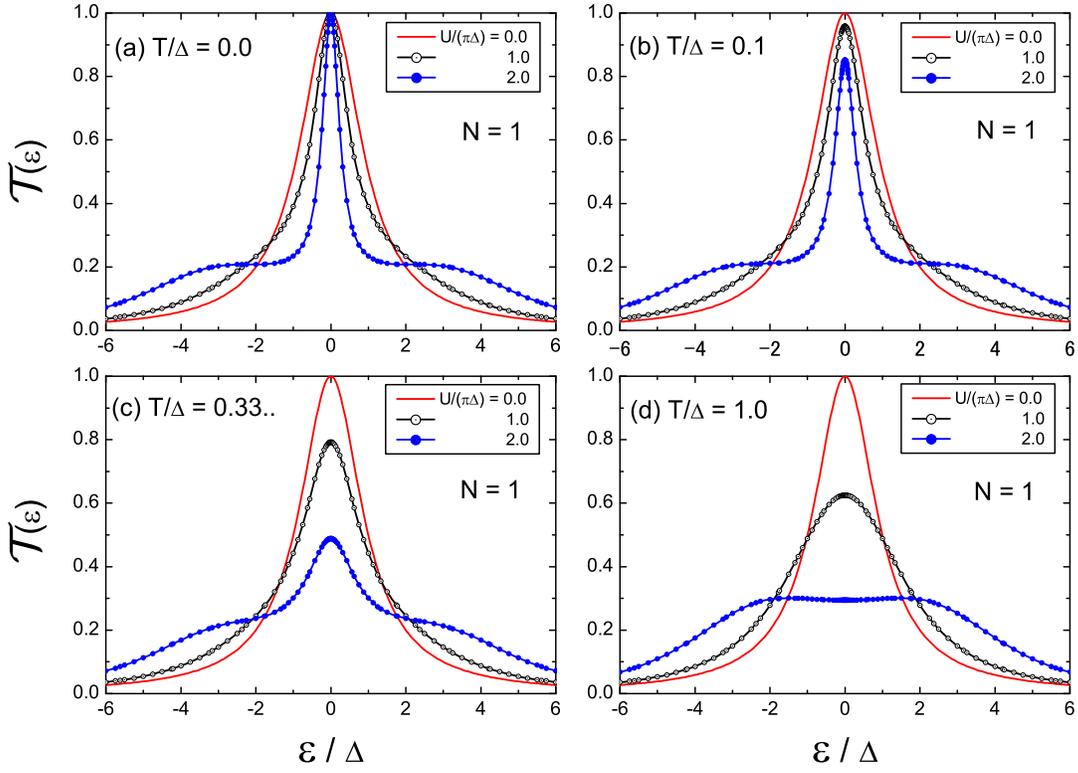}}
}
\vspace{-2cm}
\caption{Transmission probability for $N=1$. 
The temperature $T/\Delta$ is  taken to be  
(a) $0.0$, (b) $0.1$, (c) $0.333\cdots$, and (d) $1.0$.  
 The onsite interaction $U/(\pi\Delta)$ is taken to be   
(---) $0.0$, (--$\circ$--) $1.0$, and (--$\bullet$--) $2.0$.
The horizontal axis $\epsilon$ is measured 
in units of $\Delta$ ($= 2\Gamma$). 
}
\label{fig:N1}
\end{figure}

\begin{figure}
\vspace{-2cm}
\centerline{ \vbox{ \epsfxsize=160mm \epsfbox {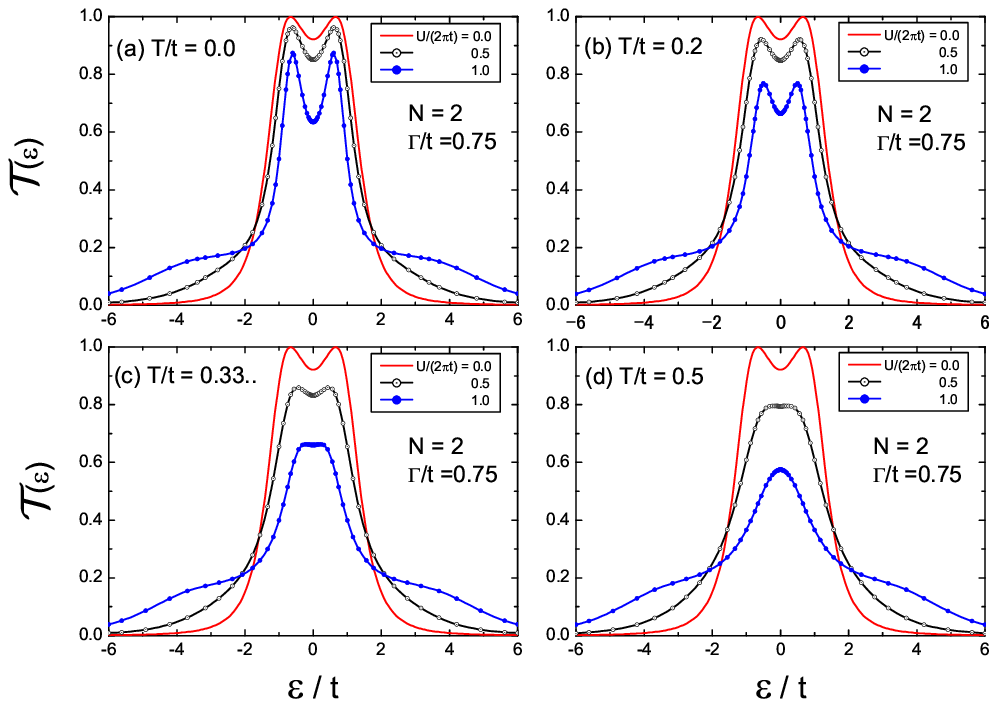}}
}
\vspace{-2cm}
\caption{
Transmission probability for $N=2$.
The temperature $T/t$ is taken to be 
(a) $0.0$, (b) $0.2$, (c) $0.333\cdots$, and  (d) $0.5$. 
The  onsite  interaction   $U/(2\pi t)$ is taken to be   
(---) $0.0$, (--$\circ$--) $0.5$, and (--$\bullet$--) $1.0$.
The horizontal axis $\epsilon$ is measured in units of $t$. 
Here $\Gamma/t =0.75$. 
}
\label{fig:N2}
\vspace{-2cm}
\end{figure}

\begin{figure}
\vspace{-2cm}
\centerline{ \vbox{ \epsfxsize=160mm \epsfbox {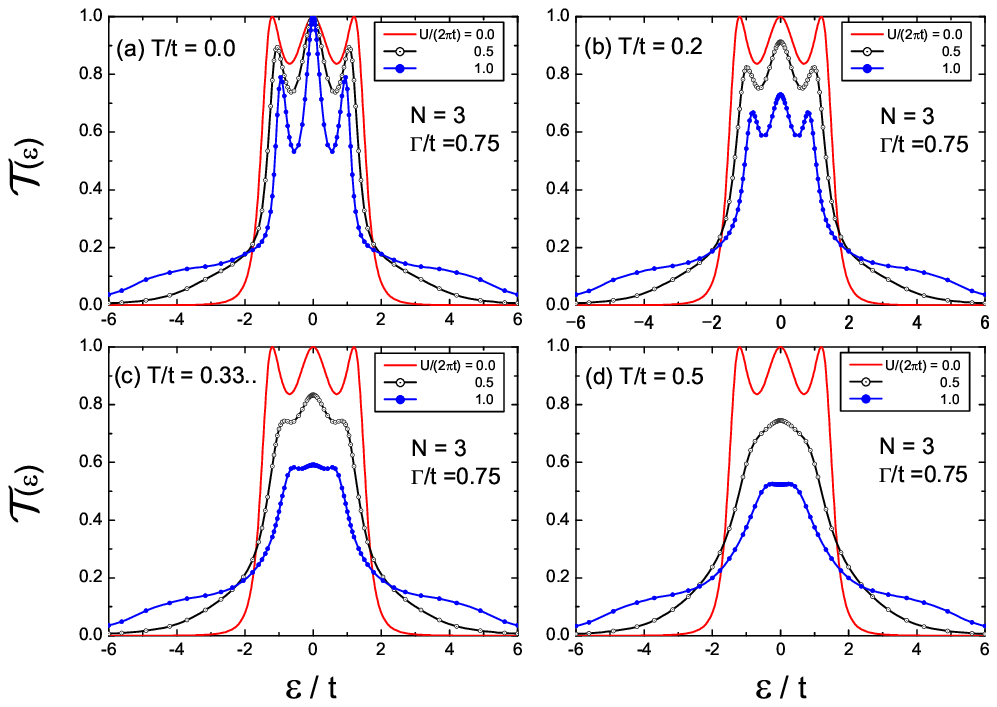}}
}
\vspace{-2cm}
\caption{
Transmission probability for $N=3$. 
The temperature $T/t$ is taken to be
(a) $0.0$, (b) $0.2$, (c) $0.333\cdots$, and  (d) $0.5$. 
The  onsite  interaction   $U/(2\pi t)$ is taken to be   
(---) $0.0$, (--$\circ$--) $0.5$, and (--$\bullet$--) $1.0$.
The horizontal axis $\epsilon$ is measured in units of $t$. 
Here  $\Gamma/t =0.75$. 
}
\label{fig:N3}
\end{figure}

\begin{figure}
\vspace{-2cm}
\centerline{ \vbox{ \epsfxsize=160mm \epsfbox {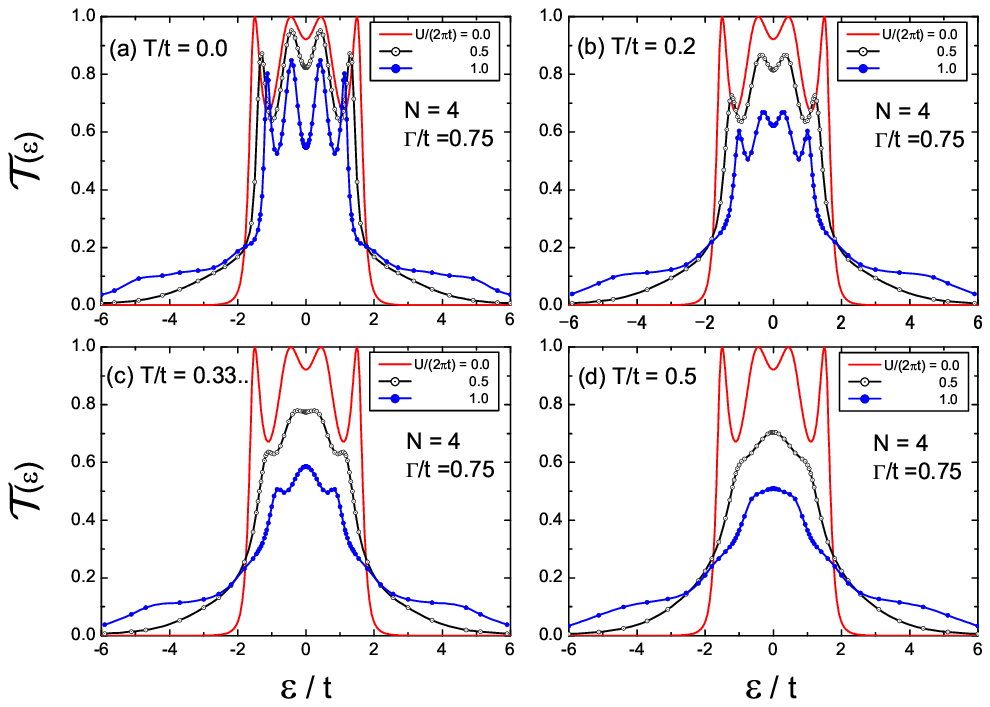}}
}
\vspace{-2cm}
\caption{
Transmission probability for $N=4$.
The temperature $T/t$ is  taken to be 
(a) $0.0$, (b) $0.2$, (c) $0.333\cdots$, and  (d) $0.5$. 
The  onsite  interaction  $U/(2\pi t)$ is taken to be   
(---) $0.0$, (--$\circ$--) $0.5$, and (--$\bullet$--) $1.0$.
The horizontal axis $\epsilon$ is measured in units of $t$. 
Here  $\Gamma/t =0.75$. 
}
\vspace{-2cm}
\label{fig:N4}
\end{figure}

We calculate the transmission probability ${\cal T}(\epsilon)$ as follows:
for the pair of the full Green's functions $G_{ij}^{\pm}(\epsilon)$ 
appearing in eqs.\ (\ref{eq:T_eff_a}) and (\ref{eq:T_eff_b})
we use the interacting ones obtained through the Dyson equation 
with the order $U^2$ self-energy, 
and for the current vertex $P^{[2]}_{j_4 j_1}(\epsilon, \epsilon)$ 
in eq.\ (\ref{eq:T_eff_b}) 
we use the order $U^2$ one, i.e., eq.\ (\ref{eq:vertex3}). 
The generalized Ward identity eq.\ (\ref{eq:Ward3_ret_0_b}) 
is satisfied in this treatment. 
We calculate 
all the matrix elements of the order $U^2$ self-energy and 
current vertex numerically 
using the unperturbed Green's function $G_{ij}^{0\pm}(\epsilon)$ 
obtained through eq.\ (\ref{eq:G_0_matrix}). 
For the reservoirs, we assume that 
the local density of states at the interfaces to be flat and 
the bandwidth to be infinity. 
Then $\mbox{\sl g}_{\alpha}^+(\epsilon)$ is 
pure imaginary and independent of the frequency: 
$v_{\alpha}^2\, \mbox{\sl g}_{\alpha}^+(\epsilon) 
= -{\rm i}\, \Gamma_{\alpha}$ for $\alpha=L, R$. 
Furthermore, we concentrate on the case 
$\Gamma_L =\Gamma_R$ ($\equiv \Gamma$), 
where the system has an inversion symmetry. 
The results of the transmission probability ${\cal T}(\epsilon)$ for 
 $N=1,2,3$, and $4$ are plotted  
in Figs.\ \ref{fig:N1}--\ref{fig:N4} for three values of $U$, 
where (a)--(d) correspond to the results at four different temperatures.
In the figures for $N=1$, the frequency $\epsilon$ is 
measured in units of $\Delta$ ($\equiv \Gamma_L + \Gamma_R$) 
which is now equal to $2 \Gamma$.
For $N \geq 2$, we measure the frequency $\epsilon$ in units
of $t$, and take the mixing parameter to be $\Gamma/t = 0.75$.

In Fig.\ \ref{fig:N1}, the transmission probability for the 
single Anderson impurity is shown for several values 
of $U/(\pi\Delta)$: 
(---) $0.0$, (--$\circ$--) $1.0$, and (--$\bullet$--) $2.0$.
The temperature $T/\Delta$ is taken to be   
(a) $0.0$, (b) $0.1$, (c) $0.33\cdots$, and (d) $1.0$. 
In this case the transmission 
probability is written in terms of the Green's function: 
${\cal T}(\epsilon) = - \Delta\, \mbox{Im}\, G^+(\epsilon)$ 
as deduced from eq.\ (\ref{eq:cond_single}).\cite{MW,HDW2} 
Therefore, Fig.\ \ref{fig:N1} corresponds to 
the results of Yamada\&Yosida\cite{YamadaYosida} and 
Horvati\'c, \v{S}ok\v{c}evi\'c \& Zlati\'c:\cite{Horvatic}  
the Kondo resonance at $\epsilon=0$ becomes sharp with increasing $U$,
and two broad peaks which have an atomic character 
appear at $\pm U/2$ for large $U/(\pi \Delta) \gtrsim 2$.
The Kondo peak decreases with increasing $T$ 
and  disappears at high temperatures in the cases of large $U$.
We note that the spectral function of 
the single Anderson impurity has been calculated accurately
with the numerical renormalization group method.\cite{CostiHewson}    
We have provided the perturbation results just for comparisons.

In Figs.\ \ref{fig:N2}--\ref{fig:N4}, 
the transmission probability for $N=2, 3, 4$ are shown 
for three values of $U/(2\pi t)$;   
(---) $0.0$, (--$\circ$--) $0.5$, and (--$\bullet$--) $1.0$.
The temperature $T/t$ is taken to be    
(a) $0.0$, (b) $0.2$, (c) $0.33\cdots$, and (d) $0.5$. 
At low temperatures ${\cal T}(\epsilon)$ for each systems 
has $N$ resonant peaks, 
which have direct correspondence with those 
of the unperturbed system.
In addition to these $N$ resonant states, 
two broad peaks appears at $\pm U/2$  in the cases of large $U$.
At $T=0$ the resonant peaks become sharp with increasing $U$, 
and valleys become deep.
However, the height of the peaks decreases with increasing $U$.
One exception is the Kondo resonance 
situated at the Fermi energy for $N=3$. 
The height of the Kondo peak is unity for any values of $U$,
and it causes a perfect transmission.
This is a general property of the odd $N$ systems, 
and occurs if the systems have the inversion 
symmetry $\Gamma_L = \Gamma_R$ together with 
the electron-hole symmetry.\cite{ao9}  
The characteristic energy scale is 
determined by the width of the Kondo peak $T_K$, 
and it decreases with increasing $N$.
On the other hand, 
the transmission probability of the even $N$ systems 
shows a minimum at $\epsilon=0$, at low temperatures. 
The characteristic energy scale is determined by 
the width of the valley which eventually tends to 
the Mott-Hubbard gap in the limit of large $N$.
Comparing the results for $N\geq 2$, 
we see that the high energy part at $|\epsilon| \gtrsim 2t$  
shows the similar slope in the case of $U/(2\pi t) =0.5$.
For this values of $U$, 
the levels of the atomic character $\pm U/2$  stay
inside the one-dimensional band edge $\pm 2t$ which 
is well-defined for large $N$.   
The two atomic levels go outside of the edge 
in the case of $U/(2\pi t) =1.0$, and 
the high energy part of ${\cal T}(\epsilon)$ for $N=3$ and $N=4$ 
show the similar behaviors at $|\epsilon| \gtrsim 4t$ 
while that for $N=2$ shows somewhat larger values.
Therefore, as far as the high-energy behaviors are concerned, 
the $N=3$ and $N=4$ systems seem to capture some aspects  
of the large $N$ systems.
The low energy part of ${\cal T}(\epsilon)$ is rather sensitive
to the temperature. 
In Fig.\ \ref{fig:N2}--\ref{fig:N4}, we see 
that the transmission probability for low-energy excitations 
at $ -2t \lesssim \epsilon \lesssim 2t$ decreases 
with increasing $T$, and 
the structures of the resonance peaks washed away 
at high temperatures.
For instance, at $T/t=0.5$ 
the transmission probability
for the $N=3$ and $N=4$ systems 
behaves almost the same  
in the whole range of $\epsilon$ in the case of $U/(2\pi t) = 1.0$. 
Thus, at high temperatures, 
the even-odd oscillation disappears.

\begin{figure}
\vspace{-2cm}
\centerline{ \vbox{ \epsfxsize=160mm \epsfbox {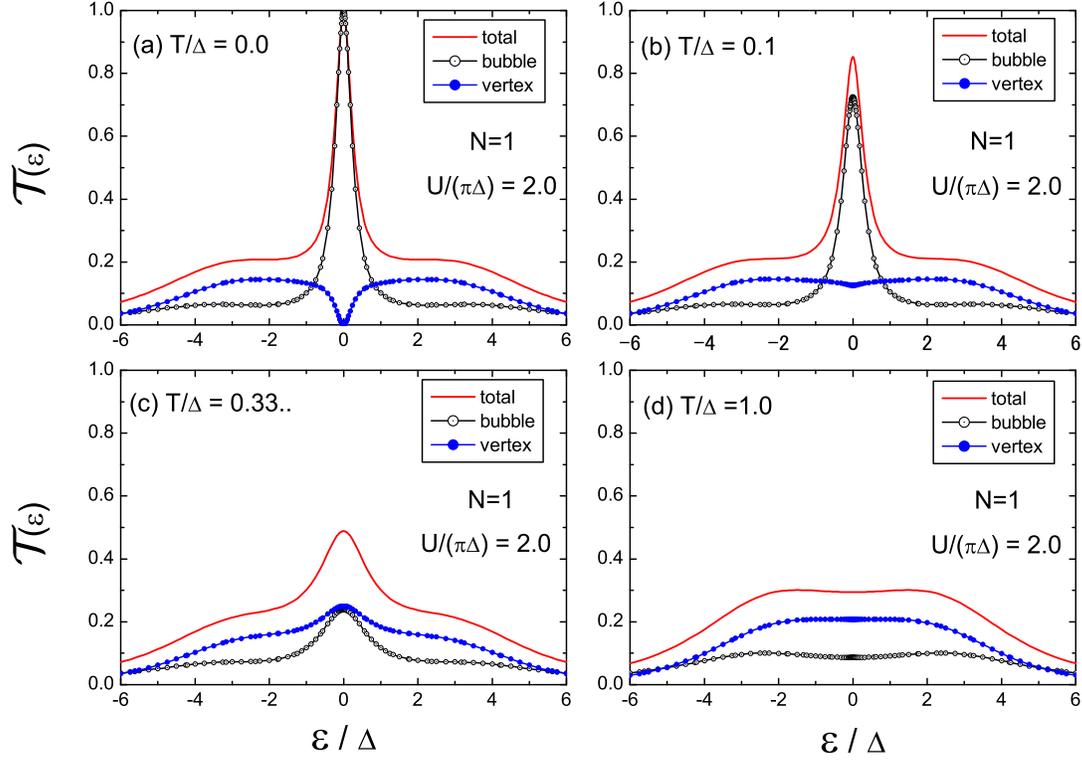}}
}
\vspace{-2cm}
\caption{
The bubble ($\circ$) and vertex ($\bullet$) 
contributions to the transmission probability (solid line) 
for $N=1$.
The temperature $T/\Delta$ is taken to be   
(a) $0.0$, (b) $0.1$, (c) $0.333\cdots$, and (d) $1.0$.  
Here $U/(\pi \Delta) =2.0$. 
}
\label{fig:N1ab}
\end{figure}

\begin{figure}
\vspace{-2cm}
\centerline{ \vbox{ \epsfxsize=160mm \epsfbox {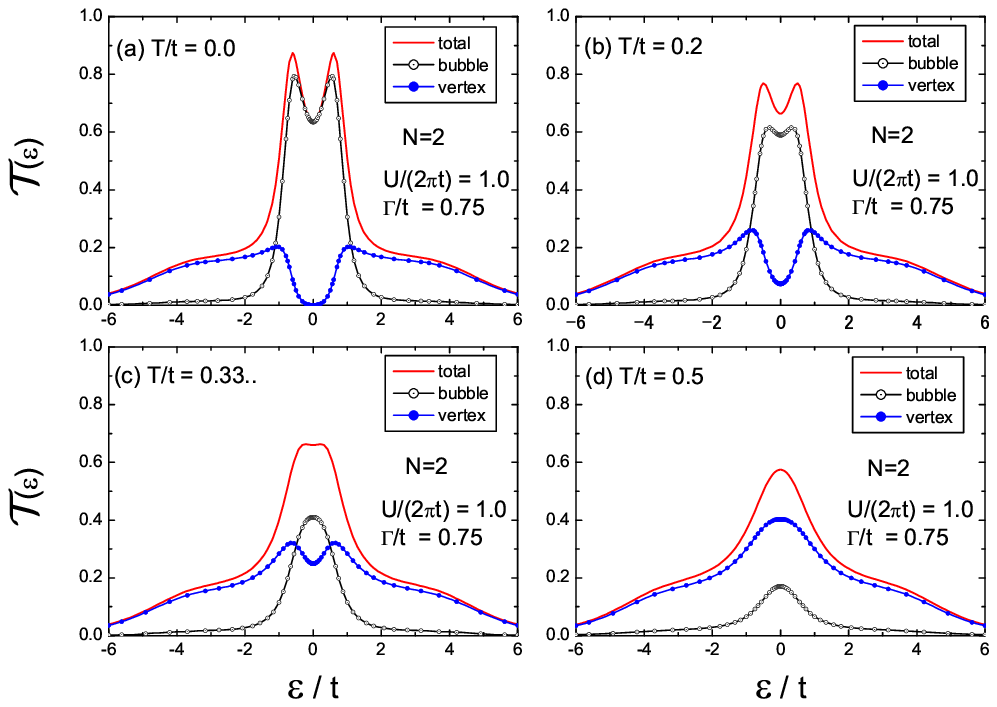}}
}
\vspace{-2cm}
\caption{
The bubble ($\circ$) and  vertex ($\bullet$) 
contributions to the transmission probability (solid line) 
for $N=2$.
The temperature $T/t$ is taken to be 
(a) $0.0$, (b) $0.2$, (c) $0.333\cdots$, and  (d) $0.5$. 
Here $U/(2\pi t) =1.0$, and $\Gamma/t =0.75$. 
}
\vspace{-4cm}
\label{fig:N2ab}
\end{figure}

\begin{figure}
\vspace{-2cm}
\centerline{ \vbox{ \epsfxsize=160mm \epsfbox {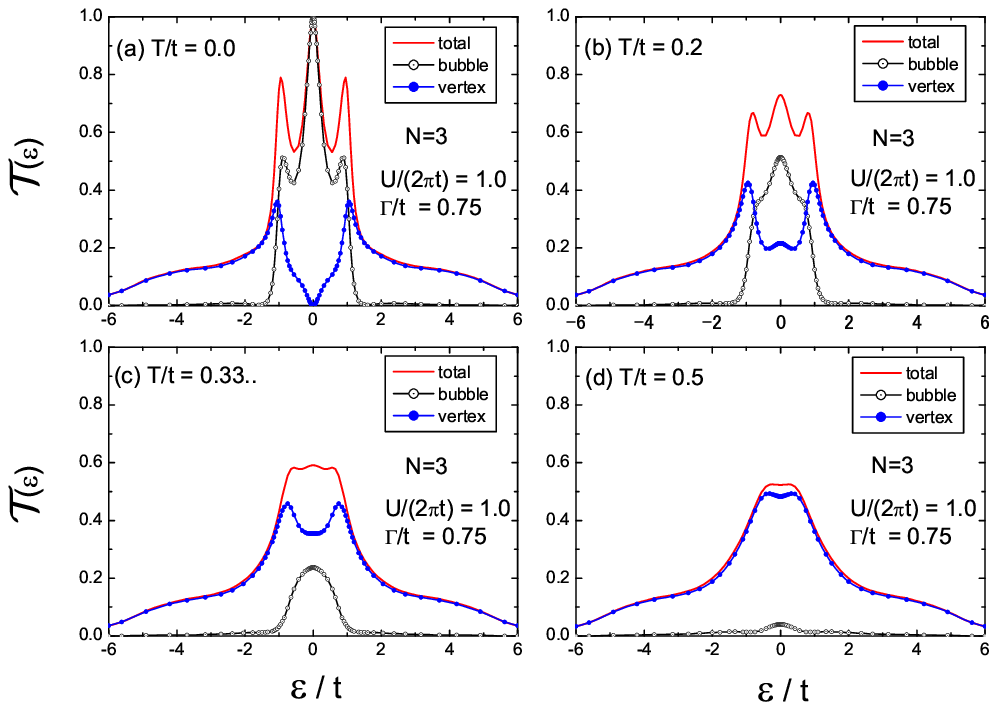}}
}
\vspace{-2cm}
\caption{
The bubble ($\circ$) and  vertex ($\bullet$) 
contributions to the transmission probability (solid line) 
for $N=3$.
The temperature $T/t$ is taken to be 
(a) $0.0$, (b) $0.2$, (c) $0.333\cdots$, and  (d) $0.5$. 
Here $U/(2\pi t) =1.0$, and $\Gamma/t =0.75$. 
}
\label{fig:N3ab}
\end{figure}

\begin{figure}
\vspace{-2cm}
\centerline{ \vbox{ \epsfxsize=160mm \epsfbox {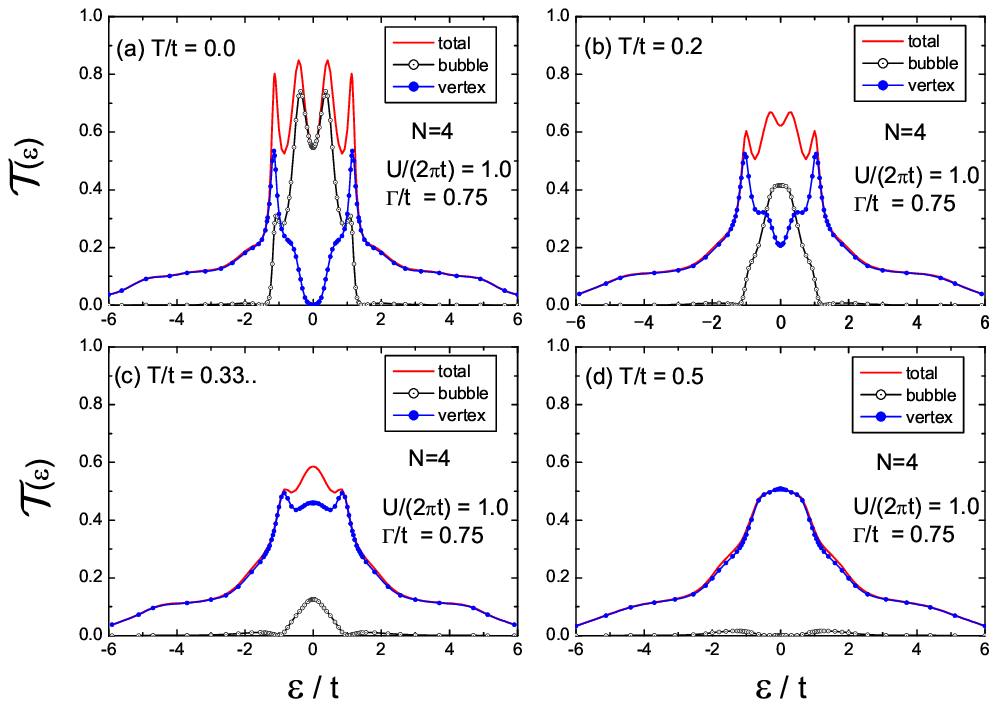}}
}
\vspace{-2cm}
\caption{
The bubble ($\circ$) and  vertex ($\bullet$) 
contributions to the transmission probability (solid line) 
for $N=4$. 
The temperature $T/t$ is taken to be 
(a) $0.0$, (b) $0.2$, (c) $0.333\cdots$, and  (d) $0.5$. 
Here $U/(2\pi t) =1.0$, and $\Gamma/t =0.75$. 
}
\vspace{-2cm}
\label{fig:N4ab}
\end{figure}

The transmission probability ${\cal T}(\epsilon)$ 
can be separated into the bubble and vertex contributions, 
i.e., ${\cal T}^{(a)}(\epsilon)$ and ${\cal T}^{(b)}(\epsilon)$
defined by eqs.\ (\ref{eq:T_eff_a}) and (\ref{eq:T_eff_b}). 
We next show how each of these two parts contributes to   
the transmission probabilities discussed above.
In Figs.\ \ref{fig:N1ab}--\ref{fig:N4ab}, 
the bubble ($\circ$) 
and vertex ($\bullet$) contributions 
are plotted for $N= 1, 2, 3$, and $4$.
In these figures,
the transmission probability ${\cal T}(\epsilon)$ 
is also shown (solid line),
and (a)--(d) corresponds to the results at different temperatures.
The value of $U$ is taken to 
be  $U/(\pi \Delta) = 2.0$ for $N = 1$, 
and $U/(2 \pi t) = 1.0$ for $N \geq 2$.
In the case of the single Anderson impurity,
${\cal T}(\epsilon)$ can be expressed in terms of 
the Green's function $G^+(\omega)$ as mentioned.
We have done the separation also in this case 
for the purpose of comparisons.
As discussed in \S \ref{sec:CONDUCTANCE}, 
the vertex contribution ${\cal T}^{(b)}(\epsilon)$ vanishes 
at $\epsilon=0$, $T=0$.\cite{ao7} 
In the present systems, ${\cal T}^{(b)}(\epsilon)$ shows 
the quadratic energy dependence for small $\epsilon$, i.e.,
${\cal T}^{(b)}(\epsilon) \propto \epsilon^2$  at $T=0$.  
This is a property of the Fermi-liquid and 
relating to the low-energy behavior of the quasi-particle 
damping, i.e.,   
$-\mbox{Im}\, \Sigma^+_{j_4 j_1}(\epsilon) \propto \epsilon^2$.
The low-energy behavior of ${\cal T}(\epsilon)$ 
is mainly determined by the babble contribution 
at low temperatures. 
However, ${\cal T}^{(a)}(\epsilon)$ decreases 
with increasing $T$.
For instance, in the $N=3$ and $N=4$ systems 
the bubble contribution almost vanishes at $T/t=0.5$ 
in the whole range of $\epsilon$, 
and the vertex contribution determines 
the total transmission probability. 
In the single impurity case, however,  
the bubble contribution remains finite even at high temperatures. 
This difference is caused by the fact that the bubble contribution 
${\cal T}^{(a)}(\epsilon) =4 \Gamma^2 |G^+_{N1}(\epsilon)|^2$
is a local quantity for $N=1$  
while it is an inter-site correlation for $N\geq 2$.
In the $N \geq 2$ systems,
the vertex contribution ${\cal T}^{(b)}(\epsilon)$ dominates 
the high energy part of the transmission probability, 
at $|\epsilon| \gtrsim 2 t$, even at low temperatures. 
The low energy part of 
${\cal T}^{(b)}(\epsilon)$ increases with the temperature. 
The value at $\epsilon=0$ shows the $T^2$ dependence at low temperatures, 
and it is also connecting with the temperature dependence 
of the damping rate $\mbox{Im}\, \Sigma^+_{j_4 j_1}(0)$ 
through eq.\ (\ref{eq:Ward3_ret_0_b}).

\begin{figure}
\centerline{ \vbox{ \epsfxsize=150mm \epsfbox {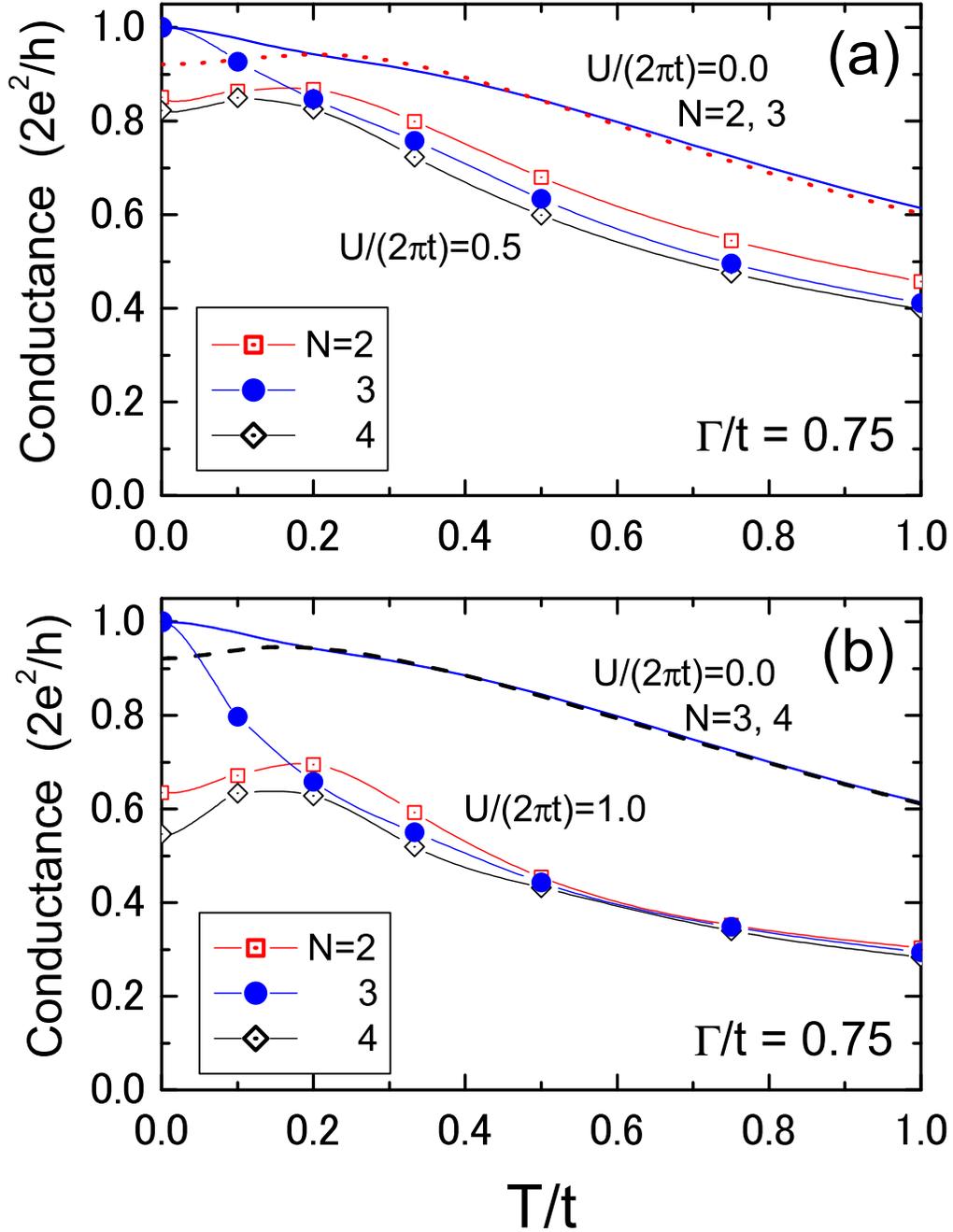}}
}
\vspace{-1cm}
\caption{ Conductance vs temperature: 
 $U/(2\pi t)$ is taken to be (a) $0.5$ and (b) $1.0$ 
for $N=2$ ($\Box$), $N=3$ ($\bullet$), and $N=4$ ($\Diamond$). 
The noninteracting results are also plotted for
$N=2$ (dotted line), $N=3$ (solid line), and $N=4$ (dashed line).
Here $\Gamma/t =0.75$. 
}
\label{fig:g_vs_T}
\end{figure}

We calculate the conductance using the results of ${\cal T}(\epsilon)$.
At low temperatures the conductance 
is determined by the low energy part of ${\cal T}(\epsilon)$ 
because of the Fermi factor $-\partial f/ \partial \epsilon$ 
in eq.\ (\ref{eq:cond1}), 
although the high energy part also 
has important information about the excitation spectrum 
as discussed in the above.
In Fig.\ \ref{fig:g_vs_T} the conductance 
is plotted as a function of the temperature $T/t$.
Here the onsite interaction $U/(2\pi t)$ is 
taken to be (a) $0.5$ and (b) $1.0$ for the systems 
of $N=2$ ($\Box$), $N=3$ ($\bullet$), and $N=4$ ($\Diamond$). 
The noninteracting results are also plotted for 
$N=2$ (dotted line), $N=3$ (solid line), and $N=4$ (dashed line).
In the case of $N=3$, the perfect transmission occurs at $T=0$, 
and the conductance decreases with increasing $T$.
This is caused by the Kondo resonance,
and is a common feature of the odd $N$ systems.
Since the Kondo peak becomes sharp with increasing $U$, 
the temperature dependence becomes steep for large $U$. 
We note that the conductance through the single impurity,
which has been obtained 
with the numerical renormalization group,\cite{Izumida3,Costi,Gerland}
shows the similar temperature dependence.
On the other hand, the conductance for even $N$ shows a minimum 
at $T=0$ because of the valley structure of ${\cal T}(\epsilon)$  
around the Fermi energy. 
Due to the reduction of the transmission probability 
at low energy part seen in Figs.\ \ref{fig:N2} and \ref{fig:N4},
the conductance decreases with increasing $U$.
Comparing the results for $N=3$ and $N=4$, 
we see that the conductance shows 
almost the same $T$ dependence at high temperatures. 
Especially in the case of $U/(2\pi t)=1.0$, 
the two curves almost overlap each other at $T/t \gtrsim 0.5$.
This is because the peak structures of ${\cal T}(\epsilon)$ are 
washed away at high temperatures 
as seen in Figs.\ \ref{fig:N3} and \ref{fig:N4}. 
This example seems to show an essential feature  
of the disappearance of the even-odd $N$ dependence.
The crossover temperature may be determined by $T_K$, 
and it decreases with increasing $N$.
The even-odd oscillation of the conductance\cite{ao9}
occurs at low temperatures $T \lesssim T_K$.

\section{Summary}
\label{sec:SUMMARY}

In summary, 
based on the Kubo formalism  
we have introduced a transmission probability ${\cal T}(\epsilon)$ for 
interacting electrons connected to noninteracting leads, 
which is given by eq.\ (\ref{eq:T_eff_sp}) and derived  
using a Eliashberg theory of the analytic continuation 
of the vertex part. 
Among the 16 analytic regions of the vertex part 
 $\Gamma^{\sigma \sigma '}_{j_1 j_2;\, j_3 j_4}
 ({\rm i}\varepsilon , {\rm i}\varepsilon '; {\rm i}\nu )$,
the transmission probability can be expressed 
in terms of the analytic function of the [2,2] region
 ${\cal L}^{\sigma \sigma' \, [22]}_{j_1 j_2;\, j_3 j_4}
(\epsilon,  \epsilon'; \omega)$. It is defined by eq.\ (\ref{eq:L22}) 
and obtained through the analytic continuation for   
${\rm i}\varepsilon  + {\rm i}\nu \to \epsilon +\omega +{\rm i}0^+$ and
${\rm i}\varepsilon  \to \epsilon -i 0^+$. 
Alternatively, ${\cal T}(\epsilon)$ can be expressed in terms 
of the three point correlation function of the current 
$\Phi^{[2]}_{R;\, j j'} (\epsilon,  \epsilon+ \omega)$ 
as eq.\ (\ref{eq:T_eff2}), and it is also written in
the Lehmann representation eq.\ (\ref{eq:Lehmann_3}).

We apply this formulation to 
a linear chain of the Anderson impurities of 
size $N$ ($=1, 2, 3, 4$) in the electron-hole symmetric case.
We calculate ${\cal T}(\epsilon)$  
using the order $U^2$ self-energy $\Sigma_{jj'} (\omega)$ and 
current vertex $P_{jj'}^{[2]}(\epsilon, \epsilon)$ 
which satisfy 
the generalized Ward identity eq.\ (\ref{eq:Ward3_ret_0_b}). 
The current conservation is fulfilled in this approximation. 
The off-diagonal (inter-site) elements  
of these functions play an important role on the transmission probability.
At low temperatures  ${\cal T}(\epsilon)$ has $N$ resonant peaks, 
which have direct correspondence with the spectrum of 
the unperturbed system.
For large $U$,  
${\cal T}(\epsilon)$ has two additional broad peaks at $\pm U/2$. 
The resonant peaks become sharp with increasing $U$. 
The peak structures are washed away at high temperatures, 
and the difference between the even and odd $N$ systems 
becomes invisible.

The formulation described in this paper 
can be generalized to multi-channel leads 
following along the similar line. 
The extension to the nonlinear response is also interesting,
and it has been done for an out of equilibrium Anderson model 
up to the third order with respect to a bias voltage.\cite{ao10}

\section*{Acknowledgements}
I would like to thank A. C. Hewson, 
H. Ishii and  S. Nonoyama for valuable discussions. 
I wish to thank the Newton Institute in Cambridge
for hospitality during my stay 
on the programme  \lq Strongly Correlated Electron Systems'.
Numerical computation in this work was partly carried out 
at Yukawa Institute Computer Facility.
This work is supported by the Grant-in-Aid 
for Scientific Research from the Ministry of Education, 
Science and Culture, Japan.

\appendix

\section{Lehmann Representation}
\label{sec:Lehmann}

In this appendix we give a Lehmann representation of 
$\Phi_{\gamma;jj'}({\rm i}\varepsilon , {\rm i}\varepsilon  + {\rm i}\nu)$
and $\Phi_{\gamma;jj'}^{[2]}(\epsilon, \epsilon + \omega)$.
Specifically, we will concentrate on the diagonal term $j=j'$,
and start with the Fourier transform;
\begin{eqnarray}
\int_0^{\beta} 
{\rm d}\tau\, {\rm d}\tau_1\, {\rm d}\tau_2\, 
{\rm e}^{{\rm i}\nu \tau} 
{\rm e}^{{\rm i}\varepsilon  \tau_1} 
{\rm e}^{-{\rm i}\varepsilon ' \tau_2}\, 
\left \langle  T_{\tau} \, J(\tau)\,
 c(\tau_1) \, c^{\dagger}(\tau_2)       
 \right \rangle  
&=& \beta\, 
    \delta_{\varepsilon+\nu,\varepsilon'}\,
    \Phi({\rm i}\varepsilon , {\rm i}\varepsilon +{\rm i}\nu)\; .
\end{eqnarray}
Here we have suppressed the subscripts to simplify the notation.
The integrations can be done explicitly 
inserting a complete set of the eigenstates, 
 ${\cal H}|n\rangle = E_n |n\rangle$, as\cite{Eliashberg}
\begin{eqnarray}
\Phi({\rm i}\varepsilon , {\rm i}\varepsilon +{\rm i}\nu)
\ &=& \  
 {1 \over Z} \sum_{lmn} \,
\langle l|c^{\dagger}|m\rangle
\langle m|J|n\rangle
\langle n|c|l\rangle 
\nonumber\\
& &\ \times
\biggl[\,
{ {\rm e}^{-\beta E_m} \over 
  ({\rm i}\varepsilon +{\rm i}\nu + E_m-E_l)({\rm i}\nu +E_m -E_n)} 
\nonumber \\  
& & \ \ \  
-\,{ {\rm e}^{-\beta E_l} \over 
  ({\rm i}\varepsilon  + E_n-E_l)({\rm i}\varepsilon +{\rm i}\nu +E_m -E_l)} 
\nonumber \\  
& & \ \ \ 
-\,{ {\rm e}^{-\beta E_n} \over 
  ({\rm i}\nu + E_m-E_n)({\rm i}\varepsilon  +E_n -E_l)} 
\,\biggr] 
\nonumber \\  
& & + \,    
 {1 \over Z} \sum_{lmn} \,
\langle l|c|n\rangle
\langle n|J|m\rangle
\langle m|c^{\dagger}|l\rangle 
\nonumber\\
& &\ \times
\biggl[\,
{ {\rm e}^{-\beta E_n} \over 
  ({\rm i}\varepsilon  + E_l-E_n)({\rm i}\nu +E_n -E_m)} 
\nonumber \\  
& & \ \ \  
+\,{ {\rm e}^{-\beta E_l} \over 
  ({\rm i}\varepsilon  + E_l-E_n)({\rm i}\varepsilon +{\rm i}\nu +E_l -E_m)} 
\nonumber \\  
& & \ \ \ 
-\,{ {\rm e}^{-\beta E_m} \over 
  ({\rm i}\varepsilon  +{\rm i}\nu+ E_l-E_m)({\rm i}\nu +E_n -E_m)} 
\,\biggr] . 
\label{eq:Lehmann_1}
\end{eqnarray}
Here $Z = \mbox{Tr}\, {\rm e}^{-\beta {\cal H}}$.
We now introduce the spectral functions defined by
\begin{eqnarray}
A_{-}(\omega_1, \omega_2) &=&
{1 \over Z} \sum_{lmn} \,
{\rm e}^{-\beta E_l}\,
\langle l|c^{\dagger}|m\rangle
\langle m|J|n\rangle
\langle n|c|l\rangle 
\, \delta(\omega_1 +E_l-E_m)
\, \delta(\omega_2 +E_l-E_n) \;,
\label{eq:A-}
\\
A_{+}(\omega_1, \omega_2) 
&=&
 {1 \over Z} \sum_{lmn} \,
{\rm e}^{-\beta E_l}\,
\langle l|c|n\rangle
\langle n|J|m\rangle
\langle m|c^{\dagger}|l\rangle 
\, \delta(\omega_1 +E_l-E_m)
\, \delta(\omega_2 +E_l-E_n) \;.
\label{eq:A+}
\end{eqnarray}
Using these spectral functions, 
eq.\ (\ref{eq:Lehmann_1}) can be rewritten as 
\begin{eqnarray}
\Phi({\rm i}\varepsilon , {\rm i}\varepsilon +{\rm i}\nu) &=&  
\,{1 \over 2} \int_{-\infty}^{\infty} {\rm d}\omega_1 {\rm d}\omega_2\,
A_{-}(\omega_1, \omega_2)\,
\nonumber
\\
& & \times
\biggl[\,
\left( 
{1 \over {\rm i}\varepsilon  + \omega_2}
+{1 \over {\rm i}\varepsilon  +{\rm i}\nu+ \omega_1}
\right)
{{\rm e}^{-\beta\omega_1} - {\rm e}^{-\beta\omega_2}  
\over
 {\rm i}\nu + \omega_1 -\omega_2 }
\,- \, 
 { {\rm e}^{-\beta\omega_1} + {\rm e}^{-\beta\omega_2} +2  
 \over
({\rm i}\varepsilon  + \omega_2)({\rm i}\varepsilon  + {\rm i}\nu+ \omega_1) }
\,\biggr]
\nonumber \\
&+&
\,{1 \over 2} \int_{-\infty}^{\infty} {\rm d}\omega_1 {\rm d}\omega_2\,
A_{+}(\omega_1, \omega_2)\,
\nonumber
\\
& &\times
\biggl[\,
\left( 
{1 \over {\rm i}\varepsilon  - \omega_2}
+{1 \over {\rm i}\varepsilon  +{\rm i}\nu- \omega_1}
\right)
{{\rm e}^{-\beta\omega_2} - {\rm e}^{-\beta\omega_1}  
\over
 {\rm i}\nu + \omega_2 -\omega_1 }
\, + \, 
 { {\rm e}^{-\beta\omega_1} + {\rm e}^{-\beta\omega_2} +2  
 \over
 ({\rm i}\varepsilon  - \omega_2)
 ({\rm i}\varepsilon  + {\rm i}\nu - \omega_1) }
\,\biggr] .
  \rule{0.5cm}{0cm}
\label{eq:Lehmann_2}
\end{eqnarray}
Carrying out the analytic continuation 
${\rm i}\varepsilon  \to \epsilon - {\rm i}0^+$ and 
${\rm i}\varepsilon  + {\rm i}\nu \to \epsilon + \omega+ {\rm i}0^+$,
and then taking the limit $\omega \to 0$, we obtain 
\begin{eqnarray}
& & \!\!\!\!\!\!\!\!\!\!\!\!\!\!\!\!
\Phi^{[2]}(\epsilon, \epsilon) \ =
\nonumber\\
& &
\, {1 \over 2} \int_{-\infty}^{\infty} {\rm d}\omega_1 {\rm d}\omega_2\,
A_{-}(\omega_1, \omega_2)\,
\nonumber 
\\
& & \times  \biggl[\,
\left( 
{1 \over \epsilon + \omega_2 -{\rm i}0^+}
+{1 \over \epsilon + \omega_1 +{\rm i}0^+}
\right)
{{\rm e}^{-\beta\omega_1} - {\rm e}^{-\beta\omega_2}  
\over
  \omega_1 -\omega_2 }
\, - \,  { {\rm e}^{-\beta\omega_1} + {\rm e}^{-\beta\omega_2} +2  
 \over
 (\epsilon + \omega_2 -{\rm i}0^+)(\epsilon+ \omega_1 +{\rm i}0^+) }
\,\biggr]
\nonumber \\
&+&
\, {1 \over 2} \int_{-\infty}^{\infty} {\rm d}\omega_1 {\rm d}\omega_2\,
A_{+}(\omega_1, \omega_2)\,
\nonumber
\\
& & \times
\biggl[\,
\left( 
{1 \over \epsilon - \omega_2 -{\rm i}0^+} 
+{1 \over \epsilon - \omega_1 +{\rm i}0^+}
\right)
{{\rm e}^{-\beta\omega_1} - {\rm e}^{-\beta\omega_2}  
\over
  \omega_1 -\omega_2 }
\,+ \, 
 { {\rm e}^{-\beta\omega_1} + {\rm e}^{-\beta\omega_2} +2  
 \over
 (\epsilon - \omega_2 -{\rm i}0^+)(\epsilon  - \omega_1 +{\rm i}0^+) }
\,\biggr] .
  \rule{0.5cm}{0cm}
\label{eq:Lehmann_3}
\end{eqnarray}
Note that  
$A_{\pm}(\omega_2, \omega_1)=A_{\pm}^*(\omega_1, \omega_2)$ 
due to the time-reversal symmetry,
and thus $\Phi^{[2]}(\epsilon, \epsilon)$ is real.

\end{document}